\documentclass[sigconf, screen=true]{acmart} % , anonymous

\AtBeginDocument{%
  \providecommand\BibTeX{{%
    \normalfont B\kern-0.5em{\scshape i\kern-0.25em b}\kern-0.8em\TeX}}}

\copyrightyear{2022}
\acmYear{2022}
\setcopyright{acmcopyright}
\acmConference[CCS '22] {Proceedings of the 2022 ACM SIGSAC Conference on Computer and Communications Security}{November 7--11, 2022}{Los Angeles, CA, USA.}
\acmBooktitle{Proceedings of the 2022 ACM SIGSAC Conference on Computer and Communications Security (CCS '22), November 7--11, 2022, Los Angeles, CA, USA}
\acmPrice{15.00}
\acmISBN{978-1-4503-9450-5/22/11}
\acmDOI{10.1145/3548606.3559388}

\usepackage{array}
\usepackage{microtype}
\usepackage{breakcites}
\usepackage{balance}
\usepackage{booktabs}
\usepackage{xcolor}
\usepackage{listings}
\usepackage{pdfcomment}

\usepackage[misc]{ifsym}

\usepackage{float}
\usepackage{stfloats}
\usepackage[flushleft]{threeparttable}
\usepackage{multirow}
\usepackage{comment}
\usepackage{enumitem}
\usepackage{seqsplit}
\usepackage[ruled,vlined,linesnumbered]{algorithm2e}
\SetAlFnt{\small}
\newtheorem{mydef}{Definition}%[section]

\makeatletter
\def\@copyrightspace{\relax}
\makeatother

\setcopyright{none}
\renewcommand\footnotetextcopyrightpermission[1]{}
\setcopyright{none}
\makeatletter
\renewcommand\@formatdoi[1]{\ignorespaces}
\makeatother

\usepackage{tikz}
\usepackage{amsmath}

\usepackage{booktabs}
\usepackage{comment}

\usepackage{xurl}

\usepackage[flushleft]{threeparttable}
\usepackage{multirow}

\usepackage{microtype}
\usepackage{breakcites}

\usepackage{booktabs}
\usepackage{xcolor}
\usepackage{listings}
\usepackage{pdfcomment}

\usepackage{float}
\usepackage{stfloats}
\usepackage[flushleft]{threeparttable}
\usepackage{multirow}
\usepackage{comment}
\usepackage{enumitem}
\usepackage{pifont}
\usepackage{makecell}

\usepackage{flushend}

\usepackage[ruled,vlined,linesnumbered]{algorithm2e}
\SetAlFnt{\small}

\def\X#1{\ding{\numexpr181+#1}}

\begin{document}

%%
%% The "title" command has an optional parameter,
%% allowing the author to define a "short title" to be used in page headers.
\title{Understanding Real-world Threats to Deep Learning Models in Android Apps}

%%
%% The "author" command and its associated commands are used to define
%% the authors and their affiliations.
%% Of note is the shared affiliation of the first two authors, and the
%% "authornote" and "authornotemark" commands
%% used to denote shared contribution to the research.

\author{Zizhuang Deng}
\affiliation{
  \institution{SKLOIS, IIE, CAS$^{\dag}$}
  \institution{School of Cyber Security, UCAS$^{\ddag}$}
\city{Beijing}
\country{China}
}
\email{dengzizhuang@iie.ac.cn}

\author{Kai Chen $^{\textrm{\Letter}}$}
\affiliation{
  \institution{SKLOIS, IIE, CAS$^{\dag}$}
  \institution{School of Cyber Security, UCAS$^{\ddag}$}
  \institution{BAAI$^{*}$}
  \city{Beijing}
  \country{China}
}
\email{chenkai@iie.ac.cn}

\author{Guozhu Meng $^{\textrm{\Letter}}$}
\affiliation{
  \institution{SKLOIS, IIE, CAS$^{\dag}$}
  \institution{School of Cyber Security, UCAS$^{\ddag}$}
  \city{Beijing}
  \country{China}
}
\email{mengguozhu@iie.ac.cn}

\author{Xiaodong Zhang}
\affiliation{
  \institution{SKLOIS, IIE, CAS$^{\dag}$}
  \institution{School of Cyber Security, UCAS$^{\ddag}$}
  \city{Beijing}
  \country{China}
}
\email{zhangxiaodong@iie.ac.cn}

\author{Ke Xu}
\affiliation{
  \institution{Huawei International Pte Ltd}
  \city{Singapore}
  \country{Singapore}
}
\email{xuke64@huawei.com}

\author{Yao Cheng}
\affiliation{
  \institution{Huawei International Pte Ltd}
  \city{Singapore}
  \country{Singapore}
}
\email{chengyao101@huawei.com}

\thanks{$\textrm{\Letter}$ Corresponding authors}

\thanks{${\dag}$ Institute of Information Engineering,  Chinese Academy of Sciences}
\thanks{${\ddag}$ University of Chinese Academy of Sciences}
\thanks{${*}$ Beijing Academy of Artificial Intelligence}

%%
%% By default, the full list of authors will be used in the page
%% headers. Often, this list is too long, and will overlap
%% other information printed in the page headers. This command allows
%% the author to define a more concise list
%% of authors' names for this purpose.
\renewcommand{\shortauthors}{}

% correct bad hyphenation here
\hyphenation{op-tical net-works semi-conduc-tor}

\newcommand\tool{AdvDroid\xspace}
\newcommand\modelset{RWM}
\newcommand\re[1]{\textcolor{black}{#1}}
\newcommand\todo[1]{\textcolor{black}{#1}}
\newcommand\dzz[1]{\textcolor{black} {#1}}
\newcommand\rev[1]{\textcolor{black} {#1}}
\newcommand\kai[1]{{\textcolor{black}{#1}}}
\newcommand\guozhu[1]{{\textcolor{black}{#1}}}
\newcommand\ke[1]{{\textcolor{black} {#1}}}
\newcommand\zxd[1]{{\textcolor{black} {#1}}}
\newcommand\cy[1]{{\textcolor{black}{#1}}}
\newcommand{\ignore}[1]{}

%%
%% The abstract is a short summary of the work to be presented in the
%% article.
\begin{abstract}
%\boldmath
Famous for its superior performance, deep learning (DL) has been popularly used within many applications, which also at the same time attracts various threats to the models. One primary threat is from adversarial attacks. 
Researchers have intensively studied this threat for several years and proposed dozens of approaches to create adversarial examples (AEs). But most of the approaches are only evaluated on limited models and datasets (e.g., MNIST, CIFAR-10). Thus, the effectiveness of attacking real-world DL models is not quite clear.
In this paper, we perform the first systematic study of adversarial attacks on real-world DNN models and provide a \underline{r}eal-\underline{w}orld \underline{m}odel dataset named \modelset{}. Particularly, we design a suite of approaches to adapt current AE generation algorithms to the diverse real-world DL models, including automatically extracting DL models from Android apps, capturing the inputs and outputs of the DL models in apps, generating AEs and validating them by observing the apps' execution. 
For black-box DL models, we design a semantic-based approach to build suitable datasets and use them for training substitute models when performing transfer-based attacks. After analyzing \dzz{245} DL models collected from \dzz{62,583} real-world apps, we have a unique opportunity to understand the gap between real-world DL models and contemporary AE generation algorithms. To our surprise, the current AE generation algorithms can only directly attack \dzz{6.53\%} of the models. Benefiting from our approach, the success rate upgrades to \dzz{47.35\%}.
\end{abstract}



%%
%% Keywords. The author(s) should pick words that accurately describe
%% the work being presented. Separate the keywords with commas.

%% A "teaser" image appears between the author and affiliation
%% information and the body of the document, and typically spans the
%% page.
% \begin{teaserfigure}
%   \includegraphics[width=\textwidth]{sampleteaser}
%   \caption{Seattle Mariners at Spring Training, 2010.}
%   \Description{Enjoying the baseball game from the third-base
%   seats. Ichiro Suzuki preparing to bat.}
%   \label{fig:teaser}
% \end{teaserfigure}

%%
%% This command processes the author and affiliation and title
%% information and builds the first part of the formatted document.

\maketitle

\section{Introduction}
\label{sec:intro}

Deep learning (DL) models, as known for their comparable or even better performance than human experts in some areas, have been widely adopted in various areas such as computer vision, object detection and speech recognition~\cite{tflite, zhao2019object, google_assist}, which also brings broader use of DL models in mobile applications (\textit{apps} for short). For example, the app PayPal~\cite{paypal} authenticates users through face recognition; the app Google Assistant~\cite{google_assist} identifies and executes users' voice commands through speech recognition. Both of the two apps have more than 10 million users with 50 million downloads. To reduce the heavy burden of computation on the server,
DL models are often designed to be stored in client-side apps. However, at the same time, the models are exposed to end-users or attackers who may leverage state-of-the-art approaches to breach the apps' defenses and finally threaten the security and privacy of users, especially when the models are undertaking security-critical tasks, e.g., authentication through face recognition and money transfer through scanning bank cards.  

Among the attacks to DL models, adversarial attacks~\cite{goodfellow2014explaining,chen2020devil,inscrypt2019} are considered as one of the most severe. It can fool the DL models through a crafted input. By adding a few perturbations on an original input (e.g., an image, a piece of audio), despite being unnoticed to humans, the new input can let a vulnerable model misclassify it to an arbitrary category. According to recent studies~\cite{ilyas2019adversarial}, deep neural networks (DNNs) are often inevitably vulnerable to adversarial attacks according to the evaluation on popular models such as VGGNet~\cite{simonyan2014very}, ResNet~\cite{he2016deep} and GCN~\cite{hao2020adversarial} using open datasets (e.g., MNIST~\cite{lecun1998gradient}, CIFAR-10~\cite{krizhevsky2009learning}, ImageNet~\cite{imagenet_cvpr09} and COCO Dataset~\cite{lin2014microsoft}). However, it is less known whether the real-world DNN models in mobile apps can be attacked. Nor do we know the real impact of successful attacks. \textbf{Thus, in this work, we aim to understand whether the DNN models in mobile apps are affected by such threats, and if so, the severity of the threats. Last, we provide the RWM dataset we collected and used in this work that could benefit researchers in understanding the limitations of current attacks and motivate them to design better attacks against real-world models.}

\vspace {3pt}\noindent\textbf{Challenges in performing real-world attacks}.
Quite different from previous research (e.g., DeepSec~\cite{ling2019deepsec}, RealSafe~\cite{dong2020benchmarking}) whose goals are often emphasized on evaluating the attack approaches, our research is end-to-end, which considers how to automatically extract DNN models from mobile apps, convert them into the forms that can be accepted by those attack approaches, generate AEs, and validate the results.
The reason for extracting models is that dynamically analyzing all the models in apps would cost a lot of time and resources.
Thus, firstly, one main challenge is to automatically extract models from mobile apps. After manually analyzing some apps, we find that the DNN models in the apps are not in fixed locations, but could be in any directories. What makes the extraction even more challenging is that the models are often with various file names, or even protected by encryption. Thus, merely searching through keywords would not work.

Secondly, even if the models are successfully extracted, they cannot be directly fed into the attacking approaches since the models' inputs and outputs, which are critical for the attacking approaches to generate reasonable AEs, are unknown. For example, in order to attack a DNN model GoogleLeNet~\cite{goodfellow2014explaining}, the input is an image of panda and the output (i.e., labels) is ``panda'', which is usually assumed to be known to attackers by default. However, when attacking real-world models in apps, such inputs/outputs are \textit{not} known. They are not simply stored in any file in mobile apps. Instead, the information about the inputs and outputs is in the semantics of the app code. For example, an app for authentication takes a photo of a user and transforms the photo into a specific format to fit the DNN model. Only by understanding the semantics of the app code, we can correctly extract the inputs and outputs. Thus, accurately identifying the code for handling inputs and outputs and understanding the code are important to successful attacks.

Thirdly, it is challenging to automatically generate and verify AEs. For white-box models, it is necessary to translate the Java code for preprocessing (e.g., normalization) in the app to Python code for AE generation. \rev{As we know, there are no such APIs in Java that allow us to perform AE attacks or even calculate gradients.} For black-box models (not using a public framework, such as Sensory~\cite{sensory}), we can perform a transfer-based attack by attacking a white-box substitute model. However, the challenge lies in building a suitable alternative dataset to train the substitute model. The alternative training dataset is required to only include the data with the same semantics indicated by the apps' output labels. 

\vspace {3pt}\noindent\textbf{Our approach}. 
To address the challenges described above, we design an approach and build a tool called \tool{} to perform large-scale end-to-end attacks on the DNN models in mobile apps, including automatically extracting the on-device DNN models, figuring out their input formats and output labels, generating AEs through various attacking approaches and validating the generated AEs, which in turn measures the performance of various AE generation algorithms. 
In particular, to address the first challenge, we find that no matter how the model's filename and location are changed, the app \todo{would} finally load the model through the APIs provided by DL frameworks. Even if the model is encrypted, the app \todo{would} decrypt the model in memory before using it for inference. Based on this observation, \rev{\tool{} performs a semantic-guided method} to extract the model file. In particular, \dzz{if the model is encrypted, \tool{} tries to automatically trigger the code that loads the model (called \textit{Model Inference Site} or \textit{MIS} for short).
After the app loads the model}, \tool{} dynamically extracts the decrypted model from memory through API hooking.

Then, \tool{} infers the inputs and outputs of the model (called \textit{model interfaces}). To achieve this goal, \tool{} first locates the MIS, then performs data flow analysis with cross references from MIS. We also design \rev{tailored} static data flow analysis to quickly determine whether there exists a path from inputs (i.e., source) to the MIS (i.e., sink), and if so, how the app preprocesses the inputs. 
Also, from the MIS, \tool{} does forward slicing to locate and recognize the outputs. 
\re{The extracted models are treated as being white-box if they can be loaded with known DL frameworks and their interface is recognized. Otherwise, it requires more steps to attack black-box models that cannot be loaded out of the app.}
\re{Note that, we focus on the models for image classification and object detection in this study, as they occupy the majority (70.31\%) of the on-devices models.}

To attack a black-box model, we leverage transfer-based attacks.  
As we know, such attacks need to train a substitute model which requires suitable inputs and labels from the target black-box model. The inputs should be meaningful to the model; otherwise, the substitute model may not be trained closely enough to the target model, which would make the attacks less successful.
To build such a dataset, we leverage the output labels extracted from the apps in the previous step. 
By comparing the semantics of the output labels in the app with the labels in the open-source datasets such as ImageNet~\cite{imagenet_cvpr09}, we can find a number of inputs with labels of similar meanings. We also leverage Google Image Search~\cite{google_image} and Open Images Dataset~\cite{open_image} to find more inputs and extend the dataset. Then we could use this dataset to train a substitute model for attacking.

\vspace {3pt}\noindent\textbf{Findings}. 
After analyzing \dzz{62,583} apps, 
we find that \re{568} apps have \dzz{960} on-device DNN models. After deduplicating the extracted models, we obtain \dzz{245} unique on-device DNN models. Among them, \dzz{177} models are white-box models which use public DL frameworks; \dzz{60} models are protected through encryption; \dzz{8} models are black-box, which means that no public DL frameworks are used. Only \dzz{16} models (\dzz{6.53\%}) of 245 models can be directly attacked by the popular attacking approaches. In contrast, using our semantic-based model interface reasoning, our approach boosts the attack success rate on model ($ASR_m$) from \dzz{6.53\%} to \dzz{47.35\% (116/245)}. Benefiting from the unique opportunity to observe the attacks on the real-world models, we have a set of interesting findings. We find that the success rates of adversarial attacks on real-world quantized models are generally 5-10\% lower than those in corresponding unquantized versions. 
\dzz{
We also find that real-world model quantization makes on-device models more robust to adversarial attacks. According to our results, the C\&W~\cite{carlini2017towards} method has the highest attack success rate on sample ($ASR_s$) among the adversarial attacks.}

\vspace {3pt}\noindent\textbf{Contributions} The contributions of this work are as follows.

\vspace {3pt}\noindent$\bullet$\space\textit{Large-scale adversarial attacks on real-world DNN models.}
We propose the first systematic study on adversarial attacks by collecting real-world DNN models and adapting them to current adversarial attacks. Particularly, we perform on-device model extraction to build a real-world model dataset RWM including \dzz{245} unique models. \rev{For each model, we build a test dataset through semantic analysis on the apps.} We also adapt the models to current adversarial attacks through a suite of new techniques, including interface reasoning (i.e., model I/O analysis) and semantic-based training data generation. The RWM dataset is released at \url{https://github.com/Advdroid/advdroid-pro}.

\vspace {3pt}\noindent$\bullet$\space\textit{New findings.} 
By analyzing the real-world models, we have the unique opportunity to understand the gap between the capability of the popularly studied adversarial attacks and the real-world situations of deployed DNN models.
We find that the real-world models are more difficult to attack (with a very low attack success rate $ASR_m$) than the commonly-used models/datasets. 
Among the attacks, the C\&W method has the best performance in attacking the real-world models. 

\section{Background}
\label{sec:bg}

\subsection{Mobile DL Frameworks}

With the growing device computational power, advanced mobile hardware accelerating techniques~\cite{chen2014diannao, chen2016eyeriss, han2016eie} and abundant RAM resource, an inference on edge devices gains its momentum nowadays.
Especially due to the increasing demand of privacy protection, on-device inference is bound to be a pivot in the near future.
Technically, DL models should be first quantized for fitting the low bit-width mobile platforms~\cite{jacob2018quantization}. The quantized models are then packed into a mobile app, e.g., an APK file. 
On Android, the models usually locate at \textit{assets} folders or exist as raw resource of varying file formats attributed to different DL frameworks. 
\re{An app may be equipped with multiple models, even developed on different frameworks, which together perform complex functions such as identifying road conditions, including traffic lights and construction zones. On the other hand, one model may be deployed on multiple apps to accomplish the same tasks. For example, developers like to use open-source models from TFHub~\cite{tfhub}.}
The model file includes the model structure and parameters, and hence there is no need to build the model in the code.
After the app is installed, the model can be loaded and run as a local module using SDK or NDK libraries of the DL framework.
Functions in code receive, pre-process, and feed the data into the local model which computes locally and returns the model output. 
  
Generally, mobile DL frameworks provide essential APIs and convenient tool sets so that developers can easily train and deploy their models on mobile devices without worrying about the detailed matrix operation or run-time optimization.
Comparing to server DL frameworks, mobile DL frameworks need to enable the model with good performance, small RAM consumption, and fast model inference.
It is more light-weight, which is usually achieved by optimized kernels, pre-fused activations, and fewer dependencies.

There are many open-source frameworks and proprietary frameworks, e.g., TensorFlow Lite (TFLite) from Google~\cite{tflite}, Caffe2 from Facebook~\cite{caffe2}.
These models, with a known format, can be invoked via public APIs. 
Therefore, it is relatively easy for an attacker to understand the model information as well as how the model is used in the app.
However, in order to protect their proprietary models, there are companies, such as DL service provider companies to whom the model is an important intellectual property, using their own proprietary mobile DL framework.
The file format of these models is unknown, and hence the model information cannot be retrieved by directly analyzing model files. 
It increases the barrier for attackers to retrieve any information from the model. 
\begin{figure*}
	\centering
	\includegraphics[width=1.7\columnwidth]{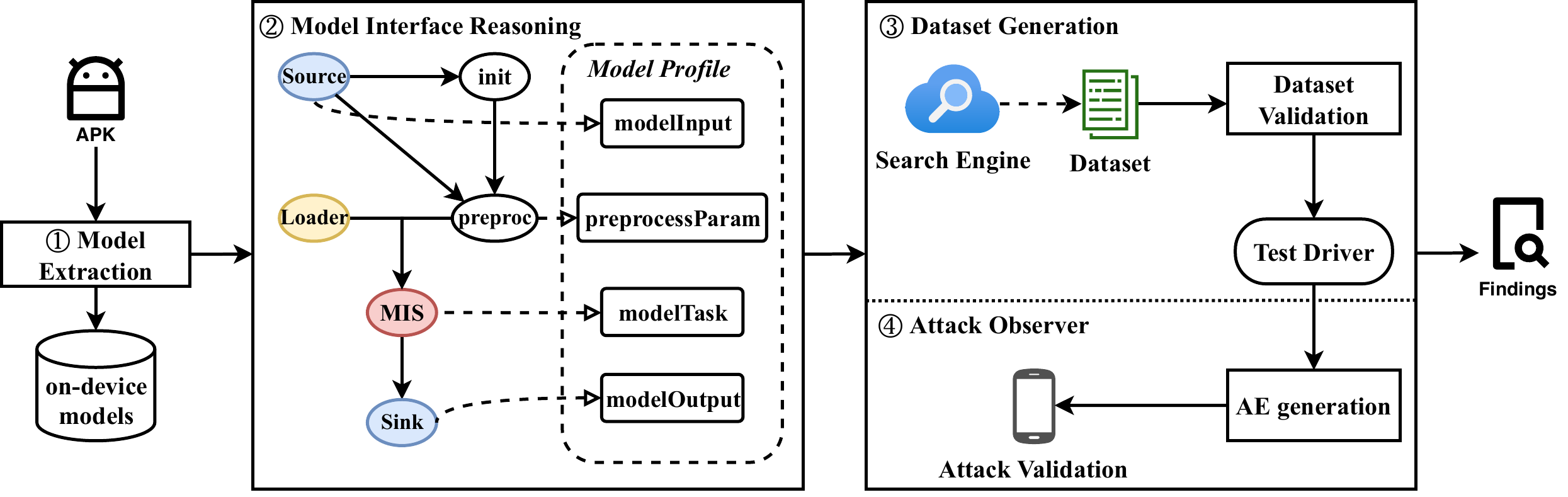}
	\caption{System overview of \tool{}}
	\label{fig:overview}
	%\vspace{-3mm}
\end{figure*}
\subsection{On-device model protection} 

It is of great importance to protect the models deployed on mobile devices. 
On one hand, as model training is expensive in both data and computational power, well-trained models are always the target of attackers.
On the other hand, the disclose of the exact model structure and parameters jeopardizes the model, as the information facilitates adversarial machine learning attacks. 
Instead of developing a proprietary mobile DL framework with unpredictable efforts, app developers are prone to using obfuscation or encryption to protect their models~\cite{sun2020mind}. 
Obfuscation is one cost-effective approach by obscuring any meaningful text in the stored model file. 
Encryption is cost-heavier than obfuscation but also provides more protection by not just obscuring meaningful text but concealing all structure and parameter information into ciphertext.

\section{Approach}\label{sec:method}

\vspace {3pt}\noindent\textbf{Threat Model.} To collect the real-world DL models for evaluating the adversarial attacks, we assume that the adversary can obtain Android apps with DL models from markets and install them on a rooted smartphone or Android emulator locally. He can freely instrument the target apps for analysis and AE generation.
In the attack scenario, we assume that the adversary can only send AEs to the victim (app) without permission to modify any environment (e.g., victim's smartphone). For example, he cannot instrument the target app.

\noindent\textbf{Overview.} 
As shown in Figure~\ref{fig:overview}, \tool proceeds in four stages: \emph{model extraction}, \emph{model interface reasoning}, \emph{dataset generation} and \emph{attack observer}.
In the model extraction module, we compile a rule list to identify the Android apps powered by deep neural networks, i.e., containing on-device models.
\re{One app may have multiple models; one model may also be deployed by multiple apps.}
By tracking the calling paths of model APIs, \tool can locate and then extract models statically. 
To cope with encrypted or packed models~\cite{sun2020mind}, \tool performs dynamic analysis to extract the plain model files.
In model interface reasoning, \tool utilizes semantic analysis to infer code semantics to obtain the information of model interfaces, including input format, model output, model task and preprocessing parameters.
In dataset generation, \tool initializes the environment for evaluating security of on-device models, which basically contains preparing drivers to load and trigger either white-box or black-box models, as well as auxiliary testing data as per model profiles.
Last, we reproduce \dzz{six} popular white-box adversarial attacks and \dzz{three} black-box ones in the module of model attack observer to reveal the models' robustness and demonstrate the caused harm.

\subsection{Model Extraction}
\label{sec:method-extract}

Given an app, \tool{} first checks whether the app contains an on-device DL model. If so, \tool{} locates the model and extracts it from the app. Below we present the details.

\vspace{2mm}
\noindent\textbf{DL App Recognition.} 
A straightforward way to check if an app contains DL models is to check the file format of every file in the app. If the file format matches the model format, we say that a model is found. As we know, the models developed on different DL frameworks have different file formats. So we investigated the top 17 DL frameworks based on market share~\cite{xu19first} and analyzed typical model formats in Table 4 %~\ref{tab:formatlist} 
in the Appendices\footnote{For the content of the appendices, please refer to the released link.}. For example, a model developed on TFLite~\cite{tflite} has the format \texttt{\small ``TFL3''}. So if any file has the format in a given app, we can extract the file as the model file. The app containing the file is a candidate DL app. 

Sometimes, apps aim to protect models by encrypting them or loading them dynamically. This makes it impossible to recognize the file format directly. To this end, we look for the code that loads the model. In most cases, the way to load models is fixed by different DL frameworks. Therefore, by identifying the code for model loading, we can ensure that the app contains the DL model. The code can also be instrumented to extract the models. We identified different file features and code features for loading models (see Table 4%~\ref{tab:formatlist}
). With these features, we can quickly identify candidate DL apps.

\ignore{
As mentioned in Section~\ref{sec:bg}, the model of one DL app may be protected through encryption, an unrecognized format, or being dynamically downloaded from the remote during runtime. As a consequence, it cannot be recognized merely based on these indicators. 
To this end, we resort to the behaviors in the app that load and use on-device models. 
We investigate the top \dzz{15} DL frameworks as per market share~\cite{xu19first} and identify the distinct features (see Table~\ref{tab:apilist} in the Appendices) in code how the corresponding models are loaded and used.
It is observed that the native DL libraries are not heavy-packed and apps that use on-device models deploy corresponding native libraries to accelerate inference. 
Based on these DL libraries, \tool{} identifies the loading code of these libraries and further determines the DL framework by summarizing the call sequences of specific APIs and involved libraries as a signature.
\rev{For example, if an app contains a native library name starting with \texttt{\small libtensorflow} and an native method named \texttt{\small NativeInterpreterWrapper.run} is called in the code, we consider this app as an DL app.}
}

\vspace{2mm}
\noindent\textbf{Model Localization and Extraction.}
After obtaining candidate apps that contain on-device models, we locate the models and extract them for further analysis.
We can extract the models directly if they are not protected. 
If a model is encrypted or dynamically loaded, we need to instrument the app for model extraction. Particularly, we divide the models into three types as follows.

\begin{itemize}[leftmargin=*]
\item \textit{Type A. Unprotected models using open-source frameworks.} 
These models are developed under open-source frameworks such as TFLite. To further verify that they can be loaded without protection, we use the loader APIs provided by the corresponding DL frameworks. For example, TFLite uses API \texttt{\small Interpreter.invoke} for model loading. If the target model can be loaded successfully, we say the model is Type-A.
\item \textit{Type B. Protected models using open-source frameworks.} 
If the model file cannot be \todo{recognized directly}, the model is very likely protected (e.g., encrypted). 
\re{To this end, \tool hooks the code for model loading (i.e., MIS), dynamically executes the app to trigger the MIS, and then dumps the model from memory after it is loaded. 
The dynamic triggering of the MIS can be facilitated via constructing a sequence of UI operations that are associated with the code of model loading.}

\item \textit{Type C. Models using closed source framework.} 
Some models may be developed with a \rev{private} DL framework, so it is difficult to load them outsides the apps since libraries or environmental requirements could not be satisfied. 
Without any details about the model, we view them as a black box and propose a dynamic approach to interact with them. More specifically, 
\tool traces API calls during running apps, selects the APIs related to model inference, and builds a remote query service with remote procedure calls (RPCs). In such a manner, we can pass crafted data to the APIs detected during runtime to obtain model output. 
\end{itemize}

Based on the above, we successfully find \dzz{960} on-device models and extract \dzz{245} distinct ones from \dzz{568} apps after deduplication \re{by hash values}. \dzz{177 (72.24\%)} models are of Type A, \dzz{60 (24.49\%)} are Type B. Besides these extracted models, \dzz{8 (3.27\%)} of on-device models belong to Type C.

\ignore{
\begin{figure}
\centering
\setlength{\belowcaptionskip}{0pt}
\includegraphics[width=0.85\columnwidth]{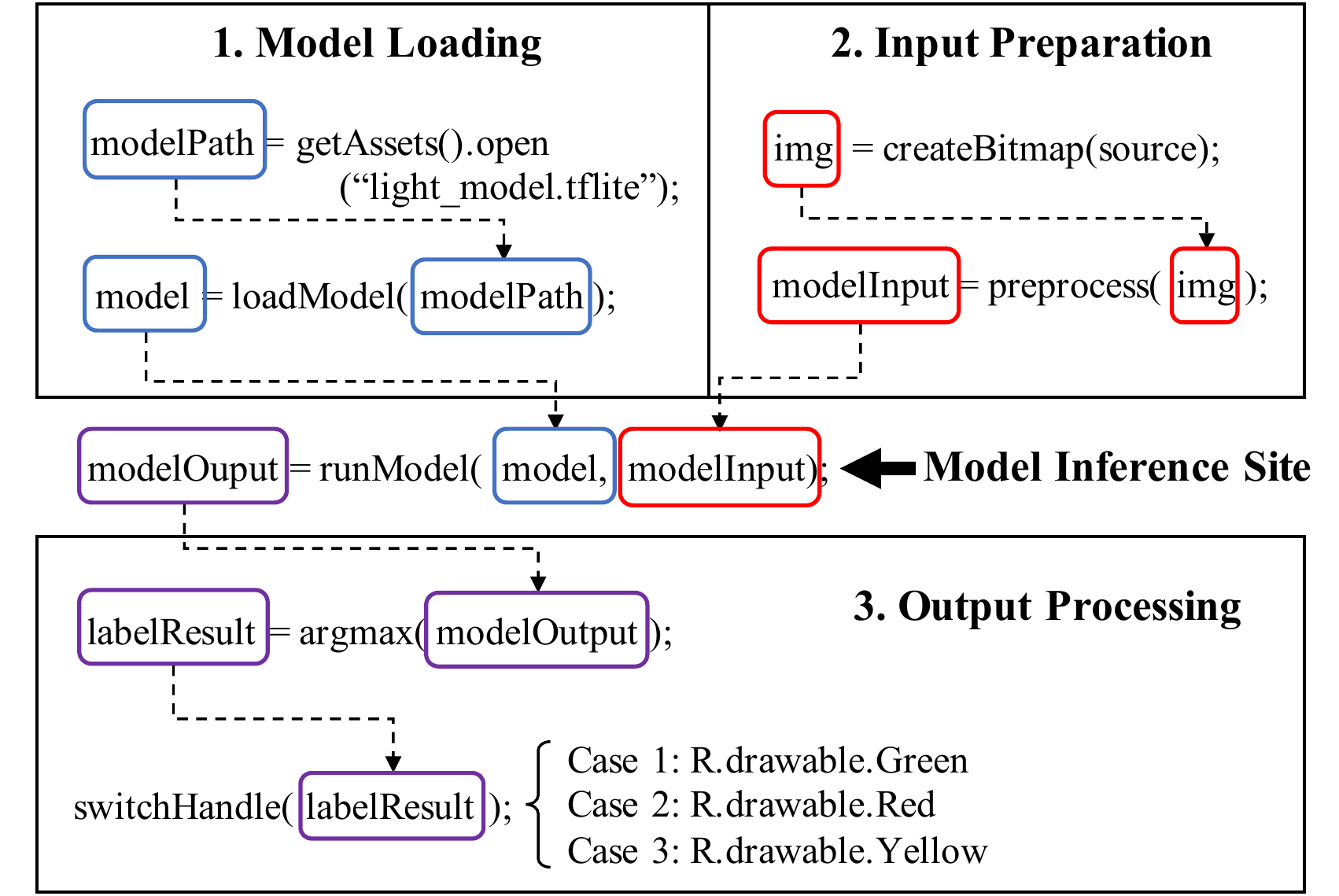}
\caption{\rev{An example of data flow with cross references of an on-device model named ``light\_model.tflite''.}}
\label{fig:dataflow}
\end{figure}
}

\begin{figure}
	\centering
	\setlength{\belowcaptionskip}{0pt}
	\includegraphics[width=0.9\columnwidth]{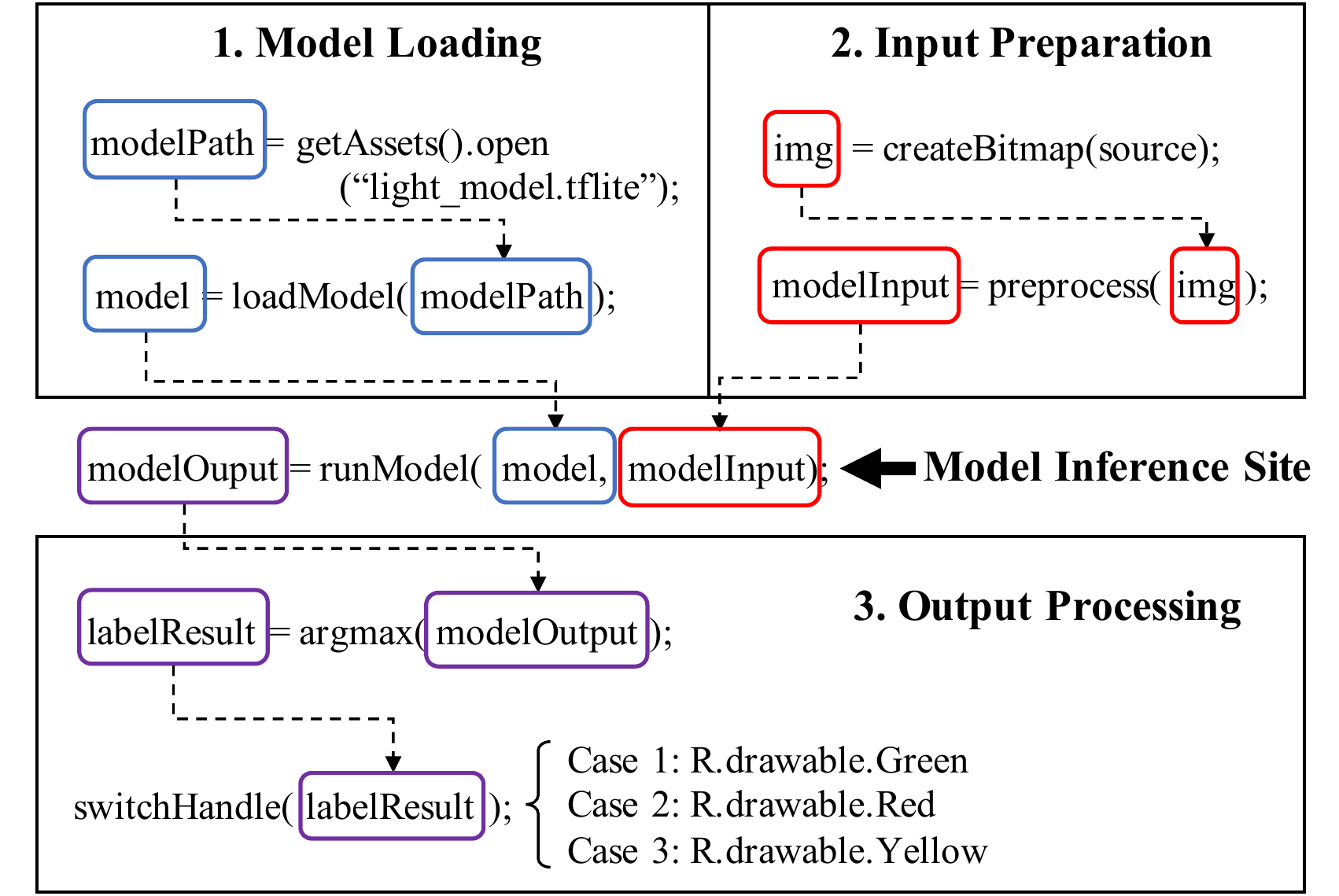}
	\caption{\rev{An example of data flow with cross references of an on-device model named ``light\_model.tflite''.}}
	\label{fig:dataflow}
	\vspace{-3mm}
\end{figure}

\subsection{Model Interface Reasoning}
\label{sec:method-interface}

\begin{comment}
\begin{table}
\scriptsize
\centering
\caption{Tag List for Model Interface Reasoning}\label{tab:infer2}
\begin{tabular}{ccc} \toprule
	\textbf{Aspect} & \textbf{Method}         & \textbf{Tag Type}         \\ 
	\midrule
	UI strings      & Resource strings        & Model task      \\
	UI strings      & Layout names            & Model scenario  \\
	Code            & Identity names          & Model interface \\
	Code            & Identity types          & Model interface \\
	Code            & Log strings             & Model task      \\
	Model           & Layer names/Attributes  & Model interface \\
	\bottomrule
\end{tabular}
\end{table}
\end{comment}

\begin{algorithm}[t]
\caption{Model Interface Reasoning}\label{alg:infer}
\KwIn{model inference site $S$, inter-component call graph $G$, and pre-defined source list $I$ and output $O$}
\KwOut{model $m$, input $i$ and output $o$}

\SetKwComment{Comment}{$\triangleright$\ }{}
$m, i, o~\leftarrow parse(S)$ \Comment*[r]{\small  initialize m, i and o from MIS} 
\While{$m \neq NULL$ }{
	$nodes \leftarrow one\_step\_backward(G,~m)$ \Comment*[r]{\small perform backward data flow analysis within one jump} 
	%$m \leftarrow backward\_flow\_analysis(G,~m)$; \\
	\If{any $n\in~nodes$ ~points~to~a~file}{
		$m \leftarrow n$ \Comment*[r]{\small find the location of model file}
		break;\\
	}{
		$m\leftarrow nodes$;\\
	}
}
\While{$i \neq NULL$}{
	$nodes \leftarrow one\_step\_backward(G,~i)$; \\
	\ForEach{$n \in nodes$}{ 
		\If{$n \in I$}{	
			$i\leftarrow n$ \Comment*[r]{\small identify the input for models}
			break;\\
		}
	}
	$i\leftarrow nodes$;
}

\While{$o \neq NULL$}{
	$nodes \leftarrow one\_step\_forward(G,~o)$; \\
	\ForEach{$n\in nodes$}{
		\If{$n \in O$}{
			$o\leftarrow n$ \Comment*[r]{\small identify the ouput for models}
			break;\\
		}
	}
	$o \leftarrow nodes$;
	
}
\Return{$m, i, o$}
\end{algorithm}

\begin{comment}
\begin{algorithm}[t]
	\caption{Model Interface Reasoning}\label{alg:infer}
	\KwIn{model name $m$ in the app, model inference $ruleList$, model loader API $f$ with input parameter $argv$ and output value $ret$ according to official documents.}
	\KwOut{Model input format, output labels, and model task.}
	$keywords\_set \gets \varnothing$;\\
	$input\_format, output\_labels, model\_task \gets \varnothing$;\\
	$start \gets find\_model\_path(m)$;\\
	$MIS \gets find\_MIS(f)$;\\
	\If{$reachability(start, MIS$) is True}{
		\rev{// If current statement met any rule in ruleList, stop slicing.} \\
		$input\_format, input\_g \gets backward\_slicing\_from\_MIS(argv)$;\\
		$output\_labels, output\_g \gets forward\_slicing\_from\_MIS(ret)$;\\
		$keywords\_set \gets \rev{tagging}(input\_g, output\_g)$;\\
	}
	\If{$input\_format$ = $\varnothing$ $\vee$ $output\_labels$ = $\varnothing$}{
		$keywords\_set \gets \rev{tagging}(start)$;\\
	}
	\rev{// Model profiling.} \\
	$model\_task \gets summarize(keywords\_set, ruleList)$;\\
	\Return{$input\_format, output\_labels, model\_task$}
\end{algorithm}
\end{comment}

To enable security evaluation of on-device models, we need to know the model's input and output.  
Figure~\ref{fig:dataflow} illustrates how one model is loaded and executed with crafted input. First, the model is loaded from a \textsf{light\_model.tflite} file via loader API ``\texttt{loadModel}''. 
Then the inference API ``\texttt{runModel}'' uses the loaded model for inference. It has two parameters: the model and the input ``\texttt{modelInput}''. Note that the input has been preprocessed (e.g., re-scaling). 
At last, the model outputs the results which are further parsed to various labels. Such input and output are important for understanding the model and also vital for further attacks. Therefore, we perform model interface reasoning to obtain the input and output.

To start with model interface reasoning, we define the Model Inference Site (MIS) as follows.
\begin{mydef}
A model inference site $\mathcal{S}$ is a concrete execution of on-device models and can be characterized as $\mathcal{S}$ = \{$m, i, o$\}, where $m$ is executed model, $i$ is the input for the model and $o$ is the output result. 
\end{mydef}

Generally, MIS associates the input and output to on-device models, which is our starting point for model interface reasoning. 
Take TFLite model in Figure~\ref{fig:dataflow} as an example. The API ``\texttt{\small runModel}'' is treated as an MIS and glues the model's input, output, and the model instance. Other DL frameworks have their own MISs, we manually summarize them in Table 5 %~\ref{tab:apilist}
in the Appendices.
From an MIS, we adopt context-sensitive data flow analysis to track the arguments and return values as stated in Algorithm~\ref{alg:infer}.
In particular, considering an MIS $(m, i, o)$, we perform backward data flow analysis from its parameter $m$ (i.e., \texttt{\small model}), which can finally locate the file ``light\_model.tflite'' (Line 2-7 in Algorithm~\ref{alg:infer}). From parameter $i$ (i.e., \texttt{\small modelInput}), we are diverted to the statement ``\texttt{\small img=createBitmap(source)}'' and learn that the model's input is an image (Line 8-14 in Algorithm~\ref{alg:infer}). 
Similarly, model output can be determined by a forward data flow analysis, where we infer the output details, e.g., pre-defined labels for a classifier (Line 15-21 in Algorithm~\ref{alg:infer}). Below we detail how to locate the model together with the corresponding inputs and outputs.

\vspace {3pt}\noindent\textbf{Model Loading}. 
Starting from an MIS, we first need to know what model $m$ it loads.
An app may contain multiple MISs and models, which should be connected correspondingly. Data flow analysis is used here to ensure the connection. For example, in Figure~\ref{fig:dataflow}, using the method proposed in Section~\ref{sec:method-extract}, we only know that the app uses the model ``light\_model.tflite''. For further analysis, we locate the MIS at the statement ``\texttt{\small runModel}''. Its parameter indicates the model is ``model''. By backtracking from the parameter, we further locate the API \texttt{\small loadModel} and identify the model as ``light\_model.tflite''.

\vspace {3pt}\noindent\textbf{Input Preparation}. We intend to infer the semantics of parameter $I$ in MIS, which is either an image, audio file or text in deep learning tasks. 
We first construct an inter-component call graph (ICCG) via class hierarchy analysis~\cite{dean1995optimization} and perform a backward data flow analysis to determine where the parameter $I$ \re{comes} from.
The analysis is terminated upon pre-defined sources are detected. These pre-defined sources are Android APIs that we have summarized in Table 6 %~\ref{tab:rules} 
in the Appendices. For example, we take methods ``\texttt{createBitmap}'' and ``\texttt{createScaledBitmap}'' as the source for creating an image, and ``\texttt{AudioRecord}'' for an audio file. 
Oftentimes, input data needs to be transformed for suiting the model. 
For example, image resizing and normalization are conventional operations before being passed to a model. 
\tool{} infers the resizing configuration from model input layer, where for example, (1, 224, 224, 3) denotes that the model accepts a three channel 224 $\times$ 224 sized image.
As for image normalization, we \re{list} a number of commonly-used APIs in image processing libraries for normalization and determine the key parameters--mean and standard deviation (see ``input preproc method'' in Table 6 in the Appendices).
Then \tool{} performs a code search in the current class to identify the possible values. 
It is worth mentioning that the analysis of input transformation is usually unnecessary for a protected model as the code is dynamically triggered with the original input.

\vspace {3pt}\noindent\textbf{Output Processing.} Here we aim to learn what results come from the model so as to prepare the necessary information for our test data. 
Take a classification task for example. We need to learn what types of objects (i.e., labels) the model can classify and then select well-suited data to feed. 
It cannot be known from viewing model structure, but instead, we can find clues from the processing code like colors in Figure~\ref{fig:dataflow}.

Similar with input reasoning, \tool identifies from the ICCG where the result generated from MIS flows to.
For example, as shown in Figure~\ref{fig:dataflow}, the app calls \texttt{switchHandle()} with model outputs as parameters and the model output is interpreted with three possible labels for a classification model.
Therefore, \texttt{switch-case} and \texttt{if-else} serve as the termination condition for our forward flow analysis. 
Besides classification, there are other types of tasks as well as outputs for a model, e.g., segmentation, style transfer and optical character recognition. 
Therefore, we summarize three types of processing handlers in terms of mainstream model tasks used in our static analysis: 
\X1 n-dimensional array of floating-point probabilities. It usually occurs in a classification task where one element of the array represents the probability of being an object. 
\X2 4-tuple of floating-point numbers. It is common in object detection to outline the boundaries of the recognized object. 
\X3 a matrix of floating-point numbers. It appears in a segmentation task where each element indicates what class the corresponding pixel is associated with. For a style transfer task, it represents an image and each element is the pixel value. 
More details about output examples can be found in Table 6.

\re{As our AE attacks focus on image classification and object detection, we retain these two kinds of models recognized as above and determine the literal labels of their output.}
Sometimes, one file like ``{labelmap.txt}'' that records the label information resides in the same folder as the model.
It is observed from the experiments that 50.31\% (\dzz{483}/\dzz{960}) of models have such files. 
Otherwise, we strip the values in \texttt{switch-case} and \texttt{if-else} statements and build the label mapping accordingly. 
Additionally, we find that \dzz{43} apps use the labels mapping of public datasets, e.g., the output value ``1'' means label ``Person'' in dataset COCO. 
If all the trials fail, \tool will use random initial samples for adversarial attacks.

\newlength{\textfloatsepsave}
\setlength{\textfloatsepsave}{\textfloatsep}

\begin{algorithm}
\caption{Dataset Validation on Model Outputs }\label{alg:validate}
\KwIn{Dataset $D=\{(X_i,Y_i)\}(0 \le i < N)$ contains top-N closest classes, $model$ with $M$ classes (M $\le$ N), threshold $\alpha_1$ and $\alpha_2$.}
\KwOut{$labelmap$.}
\SetKwComment{Comment}{$\triangleright$\ }{}
\re{// cand is a 2-dimensional list to store valid samples for each index.} \\
$cand \gets \varnothing$;  \\
$labelmap \gets \varnothing$;\\
$m \gets loader(model)$;\\
\For{$sample \in D$}{
	$model\_output \gets run\_model(m, sample)$;\\
	$index, max\_conf \gets argmax(model\_output)$;\\
	\re{// we only count the samples with high confidence ($>\alpha_1$)  } \\
	\If{$max\_conf > \alpha_1$}{
		$cand[index].append(sample's~origin~label~Y')$;\\
	}
}
\For{$index \in len(cand)$}{
    \re{// count is a temporary dictionary that stores the number of times the label appears.} \\
    $count \gets \varnothing$; \\
	\For{$item \in cand[index]$}{
		$count\{Y'\} ++$;\\
	}
	\re{// we choose the label with the most occurrences.} \\
	$sort(count)$; \\ %\Comment*[r]{\small Sorted values in descending order}
	$top\_label, top\_cnt \gets top\_1(count)$; \\
	\re{// we calculate the number of all samples in cand[index].} \\
	$tot\_cnt \gets sum(count)$; \\
	\re{// we map \textit{index} to the label that occurs most frequently. Note that the number of occurrences should reach the threshold.} \\
	\If{$\frac{top\_cnt}{tot\_cnt} > \alpha_2$}{
		$labelmap[index] \gets top\_label$;\\
	}
}
\Return{$labelmap$}
\end{algorithm}

\subsection{Dataset Generation}
\label{sec:method-setup}
Besides the model itself, we also need to form a dataset to evaluate the attacks. Since our adversarial attacks are mainly targeting classification tasks, sample inputs with the correct labels are required. Then by varying the input with different perturbation levels, we can evaluate the effectiveness of different attack methods. More inputs with various labels can better help evaluate attack methods. 

Recall that we have already known the labels of the models. So the dataset can be automatically generated by searching for the corresponding inputs using the labels. We take inputs from popular datasets (e.g., ImageNet, Microsoft COCO) and search engines. Firstly, given a model that outputs different labels, we collect inputs with the same labels from popular datasets. For example, if a label is ``cat'', \tool{} can get inputs from the open datasets. Secondly, \tool{} also uses search engines (e.g., Google image and Bing image) to obtain sufficient inputs. In addition to the image corpora, we also choose AudioSet~\cite{gemmeke2017audio} as audio corpora and Metatext~\cite{metatext} as text corpora. 

\re{However, the generated data may be not qualified for the attacks with either wrong or inaccuracy labels. For example, search engines may return a ``banana'' with the keyword ``apple'', and images associated with ``elephant'' may have refined labels like ``African elephant'' and ``tusker'' in ImageNet. Therefore, we perform data validation to get more quality data based on Algorithm~\ref{alg:validate}. 
The inputs to Algorithm~\ref{alg:validate} are top-N closest classes, the model with M classes, and two thresholds $\alpha_1$ and $\alpha_2$. The output is a $labelmap$ that maps model labels (numeric values) to strings describing the label. 
In particular, we first feed the data into the model and get the inference result (Line 5-10). Note that we only count the samples with high confidence value ($>\alpha_1$, Line 9).
\tool{} then maps a label (returned by the model) to a string. We assume that the string appearing most frequently should better describe the label (Line 11-23). Also, note that we require that the number of occurrences should reach a threshold (Line 22). 
In this way, we obtain a suitable dataset for the on-device model.}

\subsection{Attack Observer}
\label{sec:method-attack}

\tool{} provides a platform to measure the effectiveness of different attack methods against a given on-device model. The platform accepts a model as input. It also needs to know the input and output of the model. Regarding the inputs, \tool supports a template and fills the template with the resizing and normalization \todo{arguments} extracted from the app.
\todo{\tool also fills the index-label mapping into a dictionary in the template, e.g., \texttt{\small labelmap<index, label>}.}
We also define a successful attack to an on-device model. 
In particular, \tool{} \re{randomly selects 50 samples as inputs for each class from the dataset generated in Section~\ref{sec:method-setup}}. If one method can generate AEs on 80\% of them, we say that the model is defeated by this attack method~\cite{carlini2017towards}.
We implemented nine state-of-the-art adversarial attack methods (including six white-box methods and three black-box methods). 
Untargeted attack can satisfy our measurement requirements, and targeted attack is generally lower in $ASR_s$ than untargeted attack.
Details of the attack methods are shown in Table~\ref{tab:exp}. On this platform, \tool{} automatically mutates an input and generates adversarial samples with different levels of perturbation. In this way, the on-device models can be evaluated uniformly.

\section{Implementation}
\label{sec:impl}

We implement \tool with more than 5,000 lines of Python code and around 500 lines of JavaScript code.
\tool employs and extends FlowDroid~\cite{arzt2014flowdroid} to accomplish model extraction and interface reasoning.
\re{During data flow analysis, we encounter errors for 183 apps including ``timeout'', ``out-of-memory'' and ``no-sink-found''.
The average size of these failed apps is 52MB. 163 of them have multiple DEX files, and the average DEX code size is around 19MB. See Appendix E %~\ref{sec:failed} 
for details.}
By debugging and fixing these errors like considering implicit call in \texttt{\small \seqsplit{java.util.concurrent.Future}}, we successfully reduce the number of errors to 12.
For the models that cannot be directly loaded, we apply {DroidBot}~\cite{droidbot} to dynamically run them, and involve manual efforts if apps are authenticated and protected via registration. 

\todo{As for model testing, we set $\alpha_1=0.7$ and $\alpha_2=0.8$ typically for Algorithm~\ref{alg:validate}.
When performing AE attacks, \tool adapts} from projects Foolbox~\cite{rauber2017foolbox} and AdvBox~\cite{goodman2020advbox} six white-box adversarial attacks, i.e., Fast Gradient Sign Method (FGSM), Projected Gradient Descent (PGD), Deepfool, Basic Iterative Method (BIM)~\cite{kurakin2016adversarial}, Momentum Iterative Method (MIM)~\cite{dong2018boosting} and C\&W, and three black-box attacks, i.e., boundary attack~\cite{brendel2017decision}, NES attack~\cite{ilyas2018black} and substitute model transfer attack~\cite{papernot2017practical, dong2018boosting}.
\re{We use a large model ResNet152 as the substitute model following~\cite{hayes2018learning, dong2020benchmarking}. It is pre-trained on ImageNet with 75.4\% accuracy and has 60.4M parameters.}
\rev{For these models whose gradient information can be calculated, \tool uses white-box methods. Otherwise, \tool uses black-box methods, which are usually harder than white-box methods~\cite{papernot2017practical}.}
For example, although the TFLite framework is open-source, its model in our dataset lacks operators to compute gradients (see Section~\ref{sec:discuss}). \rev{It is known to be impossible so far to reverse TFLite to its Tensorflow model. It is because model quantization (Tensorflow to TFLite) uses int values (e.g., INT8) or low-precision floating points (e.g., FP16) to approximate and replace original full-precision floating weights that is irreversible~\cite{cannot_convert}. Besides, there are no official TensorFlow APIs to support the conversion from TFLite to Tensorflow.}
We then re-implement the existing transfer attack methods and build our own adversarial attack toolbox which is compatible with TFLite models.

\section{Evaluation}
\label{sec:eval}

\vspace {3pt}\noindent\textbf{App Dataset.}
We collect \dzz{62,583} apps from Google Play Store and alternative markets from May 2020 to October 2021, \re{aiming to draw more comprehensive conclusions and identify new threats and defenses over time}.
From the Google play store, we crawl the top 1,000 apps at most for all of 24 categories and obtain \dzz{22,632} apps in total. 
Similarly, we crawl the top apps from alternative markets and obtain \dzz{39,951} apps which are deduplicated by app hash values. 

\vspace {3pt}\noindent\textbf{Environment.}
The experiments are conducted on three Ubuntu 18.04 Linux servers.
One with five NVIDIA Titan X GPUs, 32 cores CPU, and 128 GB RAM is used for model attack experiments, the other two both have 128 cores CPU and 256 GB RAM for Android app analysis and on-device models extraction. 
The attack validation experiments are conducted on a Google Pixel 2 smartphone.

Here we conduct experiments to evaluate \tool and conduct comprehensive analysis of the results, in order to answer the following research questions:
\begin{enumerate}[leftmargin=*,label=\textbf{RQ\arabic*.}]
	\item How effective and efficient of \tool in extracting and analyzing on-device models? (see Section~\ref{sec:eval-eff})
	\item How are the on-device models protected from physical theft during deployment in mobile devices? (see Section~\ref{sec:system-threat})
	\item How robust of on-device models with regards to state-of-the-art adversarial attacks? (see Section~\ref{sec:ae-threat})
\end{enumerate}

\begin{table}
	\footnotesize
	\centering
	\setlength{\belowcaptionskip}{0pt}
	\caption{Distribution of different frameworks of models. Many apps use identical models so that the numbers in parentheses are counts of models after deduplication.}
	\label{tab:modelframework}
	\begin{tabular}{lrrrr}
		\toprule
		\textbf{Framework} & \textbf{Type A} & \textbf{Type B} & \textbf{Type C} & \textbf{Total} \\
		\midrule
		Tensorflow & 153 (32) & 3 (1)     & 0      & 156 (33)  \\
		TFLite     & 271 (98) & 66 (30)   & 0      & 337 (128) \\
		Caffe      & 26 (12)  & 28 (6)    & 0      & 54 (18)   \\
		Caffe2     & 6 (3)    & 13 (4)    & 0      & 19 (7)    \\
		PyTorch    & 13 (1)   & 0        & 0      & 13 (1)    \\
		PaddleLite & 33 (12)  & 21 (5)    & 0      & 54 (17)   \\
		NCNN       & 22 (10)  & 19 (9)    & 0      & 41 (19)   \\
		MNN        & 20 (7)   & 14 (3)    & 0      & 34 (10)   \\
		MindSpore  & 8 (2)    & 6 (2)     & 0      & 14 (4)    \\
		SenseTime  & 0       & 0        & 106 (2) & 106 (2)   \\
		Megvii     & 0       & 0        & 110 (2) & 110 (2)   \\
		Sensory    & 0       & 0        & 19 (1)  & 19 (1)    \\
		\textit{-Others-}     & 0       & 0        & 3 (3)   & 3 (3)     \\
		\midrule
		\textbf{Total} & 552 (177) & 170 (60) & 238 (8) & 960 (245) \\
		\bottomrule
	\end{tabular}
\end{table}

\subsection{Effectiveness and Efficiency (RQ1)}
\label{sec:eval-eff}
We evaluate the effectiveness of \tool for each phase.

\vspace {3pt}\noindent\textbf{Model extraction.}
\re{\tool{} identifies 5,573 candidate DL apps from 62,583 apps. From the 5,573 apps, \tool{} further recognizes {568} apps containing DL models, and extracts 991 models.
After a two-week manual examination, we find that 31 models are false positives. For example, some apps contain files with the extension ``*.pb''. However, the file is not a model but a ``protobuf'' file with the same file extension as models. 
So our model recognition found 960 real models and yielded a false positive rate of 3.13\%.}
Furthermore, these fake models are actually configure files: \texttt{\small feat.params} (12 occurrences), \texttt{\small METADATA.pb} (10), \texttt{\small ClientInfo.pb} (2), and others (7). 
\re{As for false negatives, we randomly sampled 100 apps from those without DL characteristics and found no DL apps.}

\vspace {3pt}\noindent\textbf{Model interface reasoning.} 
\re{We successfully identify 902 MISs, 880 input sources and 562 output labels from the extracted models.
To validate the correctness, we feed specific samples into models and examine whether models return the consistent output labels. 
The reasoning is correct and accurate if the model to test accepts the offered input and has the consistent output labels as inferred.
Based on that, it is found that 853 (96.93\%) of the input sources and 509 (90.57\%) of the output handlers are correct.}
Additionally, apps may preprocess the input for models like resizing and normalization. \tool identifies \dzz{94} such operations as well as their key parameters (e.g., mean and standard deviation in normalization).

\noindent\textbf{Dataset generation.} To further evaluate the quality of the generated data for models, we just feed them into our extracted models and the original in apps, and then obtain their inference results. 
By computing the $l_2$-norm distance of two inference results, we show how consistent of the extracted models with the original. We repeat this evaluation for 10 times. In such a manner, 475 (98.55\%) of models exhibit a high consistency under the threshold 0.1.

\noindent\textbf{\re{Runtime of \tool{}.}}
Overall, \tool{} spends about \dzz{26} hours on extracting and attacking the models in \dzz{568} DL apps. 
\dzz{On average, it takes \tool{} \dzz{9.58} minutes to attack one model in white-box attacks and \dzz{69.28} minutes in black-box attacks.}
We also evaluate the time spent on each step. 
For each app, it takes about \dzz{1.25} minutes to find and extract the on-device models. 
The time spent on inferring inputs/outputs is \dzz{5.60} minutes. 
For white-box AE generation, the average time is \dzz{2.73} minutes. 
For black-box AE generation\footnote{The iteration threshold for substitute model training is \dzz{100,000} \re{queries}.}, the average time is \dzz{62.43} minutes.

\subsection{\re{Statistics of on-device Models}}\label{sec:model_statistics}

\re{We remove the replicates from 960 extracted models via hash code and obtain 245 distinct on-device models. Here we characterize these models from the following aspects.}

\vspace {3pt}\noindent\re{\textbf{Model Frameworks \& Accessibility.}} We present the frameworks used by the models as well as their accessibility in Table~\ref{tab:modelframework}.  
It is observed that Tensorflow and TFLite contribute the most (16.25\% and 35.10\%, respectively) in the collected models.
Models with Caffe2 and NCNN frameworks are only available through the native APIs, and a higher percentage of models with these frameworks (68.42\% for Caffe2 and 46.34\% for NCNN) have lower model accessibility (i.e., Type B models).
\re{Through our experiment, we find that 170 models are protected to some extent.
Although \tool extracts 960 models, only 587 (61.15\%) of them can be loaded from an external loader with the model files, which contain 552
models of Type A and 35 
ones of Type B. For the left, i.e., 135 models of Type B and 238 ones of Type C,  we need to dynamically load and execute models in apps.}
In this study, \tool successfully runs \dzz{41} such on-device models in the apps. 

\vspace {3pt}\noindent\re{\textbf{Model Task.}} \re{Here we focus on what tasks the models can perform and the data types of model input.
As shown in Table~\ref{tab:modelusage}, the image-based models account for the largest share (70.31\%), compared to audio-based (13.13\%) and text-based (8.23\%) models.}
There are \dzz{465} image classification models, \dzz{168} object detection models including OCR models, \dzz{29} style transfer models, and \dzz{13} pose detection models among \dzz{675} image-based models.
One DL app may use several models \rev{(e.g., bank card recognition model)} while one model would be used by several DL apps \rev{(e.g., NSFW detection model)}. 
On the other hand, not all models in apps will be executed during runtime. 
For example, some SDKs (e.g., Google Firebase) have integrated on-device models for special use. Therefore, one app that relies on the SDK may not demand the DL service, and thereby will not execute the models. 

App category is associated with these models to show what types of apps are more prone to using on-device models. 
It is observed that \emph{Photo/Video} apps exhibit the most interest as object detection, optical character recognition (OCR)~\cite{googletranslate}, and VR/AR~\cite{googleAR} are the more mature areas in deep learning.
\re{After manually vetting these 245 model tasks, we find that security-critical tasks account for about 20-30\%. 
Top security-critical tasks include face recognition (23), identity card recognition (14), road condition recognition (3), malware detection (2), and so on.}
The models in Finance category are mainly used for face recognition, and the models in category \emph{Communication} are mainly used for speech recognition.
Besides, the apps in \emph{Education} usually use models for OCR and keyboard typing, and the apps in \emph{Tools} use models usually for translation to provide accessibility service.

\vspace {3pt}\noindent\textbf{Model Optimization.}
In order to reduce the size of models and raise inference efficiency, most mobile deep learning frameworks quantify numerical representation in parameters and reduce fused operators used in the training process.
\re{In the 587 loadable models that can be peeked into for network internals, there are 261 models with optimization characteristics, including quantization (238) and pruning (23). 
Additionally, there are 200 (84.03\%) quantized models using the TFLite framework, 
196 quantized models use the 8-bit uint quantization method, and the rest 42 models use 16-bit uint quantization method and else. 
Quantization parameters can be directly obtained from model files, for example, by calling TFLite API \texttt{\seqsplit{interpreter.get\_input\_details()[0][`quantization']}}.}
The most used quantized parameters are $(zero\_point=0.0078125, scale=128)$.
We consider that a model pruning exists if the ratio of zero weights and biases is \re{larger} than a certain threshold (e.g., 40\%~\cite{liu2018rethinking}).
As such, we find that \dzz{23} models are pruned.

\section{Measurement and Findings}
\label{sec:measure}

\begin{table}
	\footnotesize
	\centering
	\caption{Model input types in apps of different categories}
	\label{tab:modelusage}
	%\begin{tabular}{l|l|l|l|l||l}
	\begin{tabular}{lrrrrr} 
		\toprule
		\textbf{Category} & \textbf{Image} & \textbf{Audio} & \textbf{Text} & \textbf{UNKNOWN}  & \textbf{Total} \\ 
		\midrule
		Photo/Video   & 274     & 34    & 2    & 12      & 322   \\ 
		Entertainment & 95      & 20    & 6    & 14      & 135   \\
		Beauty        & 87      & 32    & 3    & 8       & 130   \\ 
		Tools         & 27      & 10    & 21   & 14      & 72    \\
		Finance       & 43      & 2     & 10   & 0       & 55    \\ 
		Communication & 32      & 5     & 2    & 9       & 48    \\
		Education     & 23      & 4     & 19   & 1       & 47    \\ 
		Medical       & 12      & 0     & 3    & 1       & 16    \\ 
		\textit{-Others-}        & 82      & 19    & 13   & 21      & 166   \\
		\midrule
		\textbf{Total}& 675     & 126   & 79   & 80     & \rev{960}   \\
		\bottomrule
	\end{tabular}
	%\vspace{-5mm}
\end{table}

In this section, we conduct an empirical analysis of physical theft threat and evaluate the robustness of on-device models under adversarial attacks.

\subsection{Threats of Physical Theft (RQ2)}
\label{sec:system-threat}

Intellectual property infringement becomes a severe threat for deep learning models and has yet to be solved~\cite{tse2020dlsurvey}. 
Prior research has paved a way to unveil the possibilities of stealing models through authorized APIs~\cite{usenix2021drmi,usenix2021watermarks}. 
However, this threat is significantly worse when offline models are deployed in mobile devices where the models may be physically accessed and stolen by attackers. 
It shows that on-device models have already performed limited measures to protect their weights~\cite{sun2020mind}.
\re{In this section, we investigate the host apps for the extracted 960 models, introduce the existing protection measures that are being used in reality and unveil the flaws threatening these models.}
Below we compile a list of model protection techniques from two perspectives.

\vspace {3pt}\noindent\textbf{System-level protection.} Apps treat on-device models as a sensitive component and protect them in conventional manners. 
\begin{itemize}[leftmargin=*]
	\item Remote loading. Models are not stored in apps so that static analysis and code viewing cannot find them. Otherwise, models are dynamically loaded from the remote during runtime and perform inference for one time. As observed, there are 4 apps that behave in such a way in our dataset. 
	For example, we find that one app contains a DL native library and MIS ``\texttt{\small MgFaceLab::Miguface}'', but the size of the associated model file ``\texttt{\small thin.tflite}'' is zero. 
	By monitoring its execution, we identify that the model content comes from HTTP requests, and the model is used to extract features of face images and then upload them to one server for further detection and recognition.
	
	\item Model encryption. Alike other key components, on-device models are often encrypted and the ciphertext is stored locally. Apps need to decrypt ciphertext and load the model in memory for use. We find that \todo{23} apps have taken this measure to protect their models. \todo{The popular encryption methods include simple bitwise computation (5), TEA~\cite{tea} (2), AES (10), etc.}
	\item Model packing. As described by the document of MACE framework~\cite{modelpacking}, models can be converted to native C++ code that is much harder for parsing. That is, one model is stored as native code rather than a plain file. \todo{It is common in reality and 16 models cannot be extracted.}
	\item Identity authentication. Models are only granted to specific users for use. Therefore, some apps authenticate whether the current user owns a valid authorization token. The token is generally distributed during app execution by a remote server for the registered users. For example, we find that an app requires users to register first to use the model for liveness detection.
	\item Integrity verification. It is used to verify whether models are stolen and apps are repackaged. More specifically, models may be replaced with a poisoned one, or code is changed to bypass authentication by attackers. To handle this problem, apps can compute the hash code for the model and app itself in advance. On-device inference is only conducted if integrity is verified successfully. We find 3 cases that have verified model integrity.
\end{itemize}

\vspace {3pt}\noindent\textbf{Model security enhancement.} System-level protections can be hacked by mature techniques like instrumentation, debugging, and simulation. It is intriguing to explore the measures to enhance the security of on-device models from physical theft. In this study, we identify the following three out-of-the-box techniques.
\begin{itemize}[leftmargin=*]
	\item Layers obfuscation. Layers are commonly assigned with lucid names in model design. For example, ``Conv2D'' is a 2D convolution layer while  ``AvgPool2D'' implies a layer of performing average pooling. We find some apps obfuscating model layer name in our dataset, e.g., \seqsplit{``7cff058686c711e9a0ac4ccc6ac78afa:2''}. Although ineffective in protecting models from theft and misuse, it can increase the difficulty of model interpretation and further model development.
	\item Weights transformation and protection. Weights are vital for a model and should be protected as well. Prior studies have proposed several methods, e.g., placing parts of the model in TEE~\cite{bayerl2020offline} 
	and masking model weights during deployment and unmasking them during runtime~\cite{sun2020shadownet}. Although we find no apps with this type of protection, it exhibits a superior efficacy in model protection and will be widely applied once the additional overhead is significantly reduced.
	\item Custom operators. Model developers may create their own operators with DL frameworks, seeking for more flexible and powerful computations beyond built-in operators (e.g., convolution, relu, pooling, normalization). These operators, in the meantime, raise the difficulty of misusing the models. Unless the attackers learn how operators compute, they cannot load and use these models. In our dataset, we find 20 models with custom operators that are defined in native libraries. For example, in app Google AR~\cite{googleAR}, an operator ``\texttt{\small MaxPoolingWithArgmax2D}'' is designed to perform max pooling and output the pooling indices. 
\end{itemize}

\re{\textit{
Through an analysis of these protection techniques, we identify some development flaws, as described in Table 7  
in the appendices.
We have reported these issues to the affected vendors and received one confirmation. We also find that \todo{four} issues are fixed in the newer versions.
As shown in Table 7
, we do find an app that applies more defenses to protect models in its new version. But it does not necessarily imply that on-device models become more secure over time.
We also provide some suggestions in Section~\ref{sec:guide}.}}

\begin{table*}[ht]
\footnotesize
\centering
\caption{Untargeted attack results with $\epsilon=0.06$ (16/255) under the $l_\infty$ norm}
\label{tab:exp}
\begin{threeparttable}
\begin{tabular}{c|c|c|c|c|c|c|c|c|c|c|c|c}
%\toprule
\Xhline{2\arrayrulewidth}

\multirow{2}{*}{\textbf{Model Type}} & \multirow{2}{*}{\textbf{Test Difficulty}} & 
\multirow{2}{*}{\textbf{Num\tnote{1}}} & 
\multicolumn{6}{c|}{\textbf{White-box method}} & 
\multicolumn{3}{c|}{\textbf{Black-box method}} &
\multirow{2}{*}{\textbf{Succ Num\tnote{1}}} \\
\cline{4-12}
 & & & \textbf{FGSM} & \textbf{PGD} & \textbf{Deepfool} & \textbf{BIM} &\textbf{MIM} & \textbf{C\&W} & \textbf{NES} & \textbf{Boundary} & \textbf{Transfer} &  \\
\cline{1-13}
\multirow{2}{*}{A} & direct test     & 21   & 14  & 16  & 14  & 13  & 15  & 14  & 9  & 8  & /  & 16 \\
\cline{2-13}                                        
  & interface adaption               & 156  & 63  & 52  & 56  & 51  & 45  & 66  & 22 & 12 & /  & 80 \\
\cline{1-13}
\multirow{2}{*}{B} & dynamic extraction & 27  & 3  & 6  & 5  & 5  & 5  & 6  & 5  & 3  & 5  & 7 \\
\cline{2-13}
  & black-box query                  & 33  & /  & /  & /  & /  & /  & /  & 4  & 9  & 4  & 9 \\
\cline{1-13}
C & black-box query                  & 8   & /  & /  & /  & /  & /  & /  & 3  & 1  & 3  & 4 \\
\cline{1-13}
Total &                              & 245 & 80 & 74 & 75 & 69 & 65 & 86 & 43 & 33 & 12 & 116 \\

\Xhline{2\arrayrulewidth}
\end{tabular}
\begin{tablenotes}
\item [1] ``Num'' indicates the number of models, \todo{``Succ Num'' indicates the number of models successfully attacked by any one method.}

\end{tablenotes}
\end{threeparttable}
\end{table*}

\begin{figure}[!ht]
	\centering
	\includegraphics[width=0.4\textwidth]{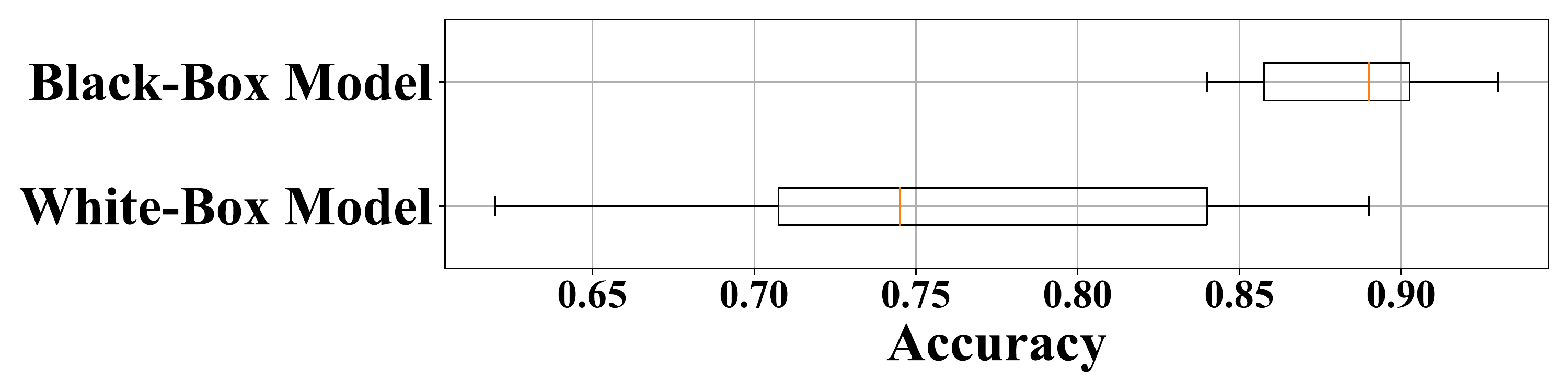}
	\caption{The accuracy of different models tested on our generated test dataset.}
	\label{fig:model_accuracy}
	\vspace{-5mm}
\end{figure}

\subsection{Threats from Adversarial Attacks (RQ3)}
\label{sec:ae-threat}
In this section, we explore how robust on-device models considering adversarial examples. 
\re{Note that, we only test image-based models since the models of other types are too few to draw convincing conclusions. Even so, our attacking approach can be easily adapted to them by employing proper AE generation algorithms.} 

\vspace {3pt}\noindent\textbf{Test dataset.}
For each model to test, we prepare a dataset for adversarial attacks from public datasets (e.g., ImageNet~\cite{imagenet_cvpr09}, Open Images~\cite{open_image}) and Google Image. 
As such, we obtain \dzz{46} datasets in total for these models that contain \dzz{251,765} samples. 
We manually evaluate the accuracy of semantic-based training data generation.
In total, \dzz{20} datasets are generated with \dzz{320} labels during white-box attacks, and \dzz{26} datasets are generated with \dzz{620} labels during black-box attacks. The accuracy of 245 models tested on our datasets is shown in Figure~\ref{fig:model_accuracy}.

\vspace {3pt}\noindent\textbf{Test difficulties.}
Models are varying from test difficulties considering protection measures. Here we define four types of test difficulties to launch an adversarial attack. 
\begin{itemize}[leftmargin=*]
	\item Direct test means that we can run adversarial attacks to evaluate models with no adaptions. 
	\item Interface adaption denotes that it needs to provide any input preparation to adapt the models, like normalizing input images.
	\item Dynamic extraction means that the plain models need to be extracted in advance for loading outside the app.
	\item \re{Black-box query necessitates an app trigger to dynamically send input samples to the model and then get the attack results, amounting to 41 models (33 Type-B and 8 Type-C models).} 
\end{itemize}

As shown in Table~\ref{tab:exp}, \dzz{116} of 245 tested models can be successfully attacked by AEs. 
\re{Except 16 Type A models of ``direct test'' difficulty, 100 (86.21\%) of them are benefited from interface reasoning.} 
We also find that white-box attack methods can attack more models than black-box methods.

\begin{figure}
	\centering
	\includegraphics[width=0.4\textwidth]{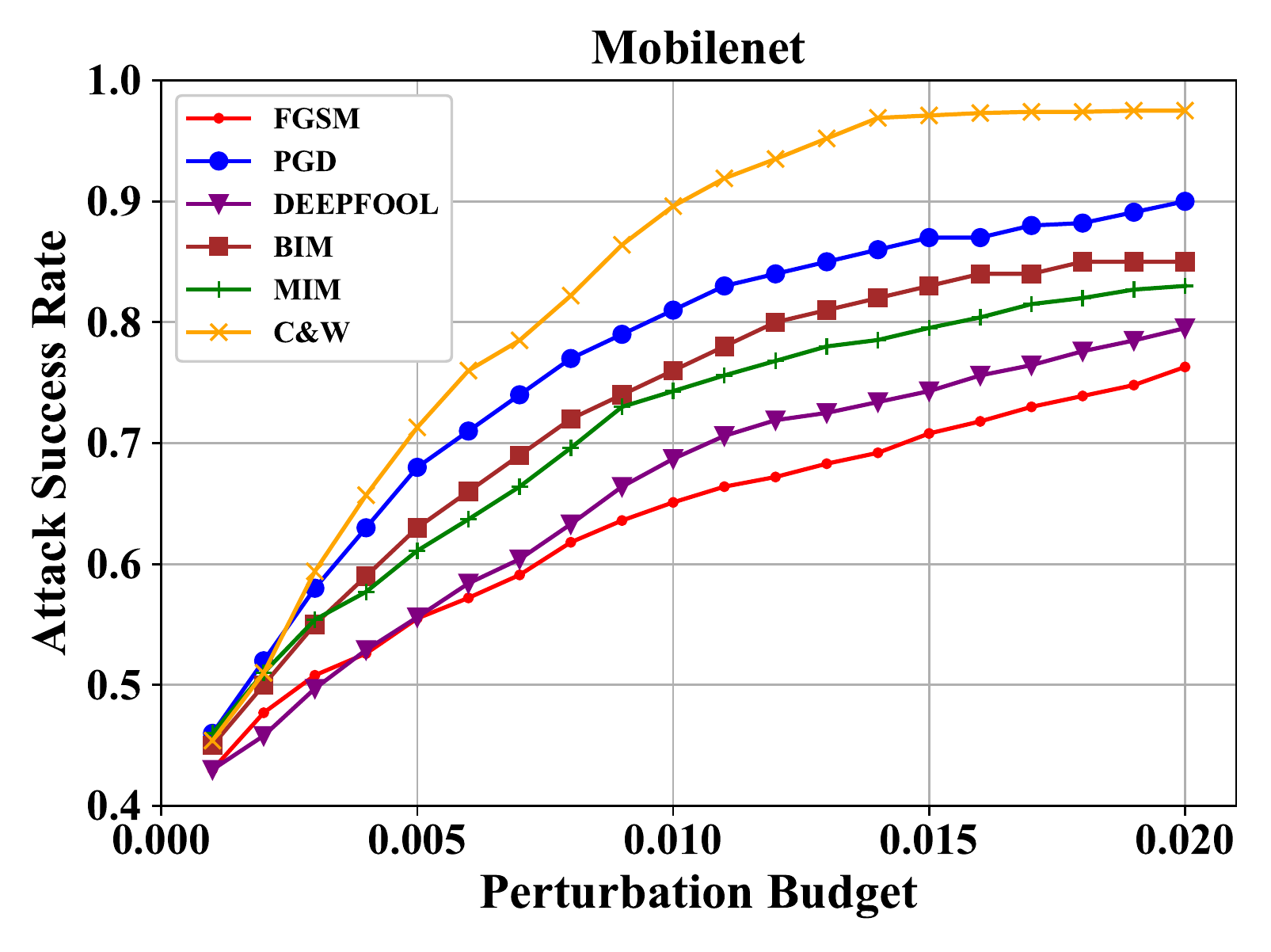}
	\caption{Performance comparison on the most-used real-world model MobileNet against untargeted white-box algorithms under $l_\infty$ norm.}
	\label{fig:diff_algo}
	\vspace{-5mm}
\end{figure}

\begin{figure}
	\centering
\includegraphics[width=0.4\textwidth]{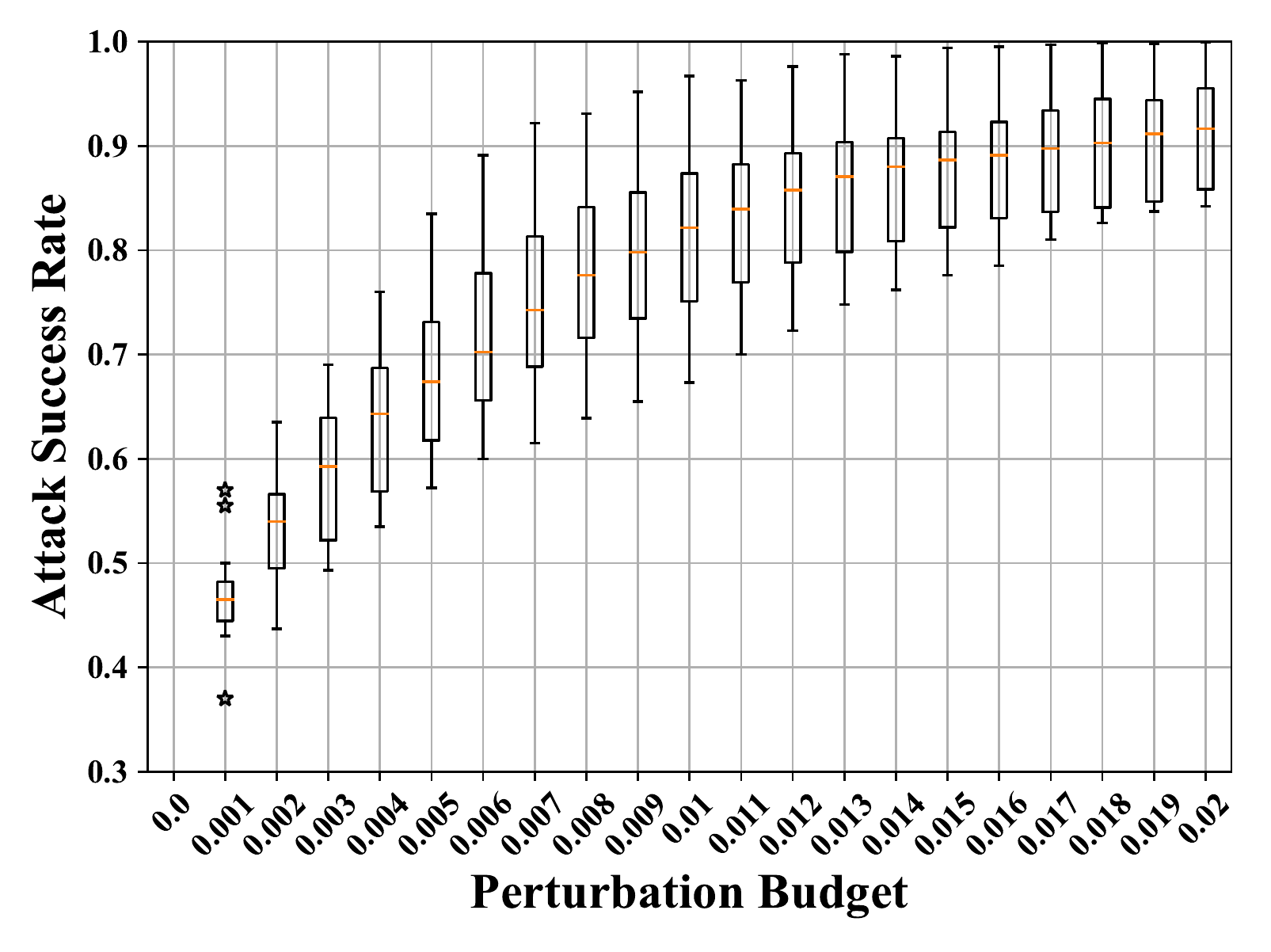}
	\caption{Distribution of $ASR_s$ of white-box models over the $l_\infty$ perturbation budgets in untargeted FGSM attack.}
	\label{fig:mult_boxplot}
	\vspace{-5mm}
\end{figure}

\vspace {3pt}\noindent\textbf{White-box adversarial attacks.}
For white-box models, \tool{} performs \dzz{six} different white-box attacks. The results are shown in Table~\ref{tab:exp}. In terms of the number of successful attacks, i.e., $ASR_m$, C\&W has the best performance.
\todo{For the FGSM method, we draw a box plot in Figure~\ref{fig:mult_boxplot} to show the distribution of $ASR_s$ of white-box models with different perturbation budgets.
We find that the distribution of $ASR_s$ of different models gradually diverges as perturbation budget increases. 
Additionally, the models with high $ASR_s$ are generally those with a large number of classes (e.g., > 10) and complicated model structures (e.g., over 100 model layers and 20,000 model neurons). Thus, providing real-world models is of great value to security researchers.
}

\begin{figure*}
	\centering
	\includegraphics[width=0.75\textwidth]{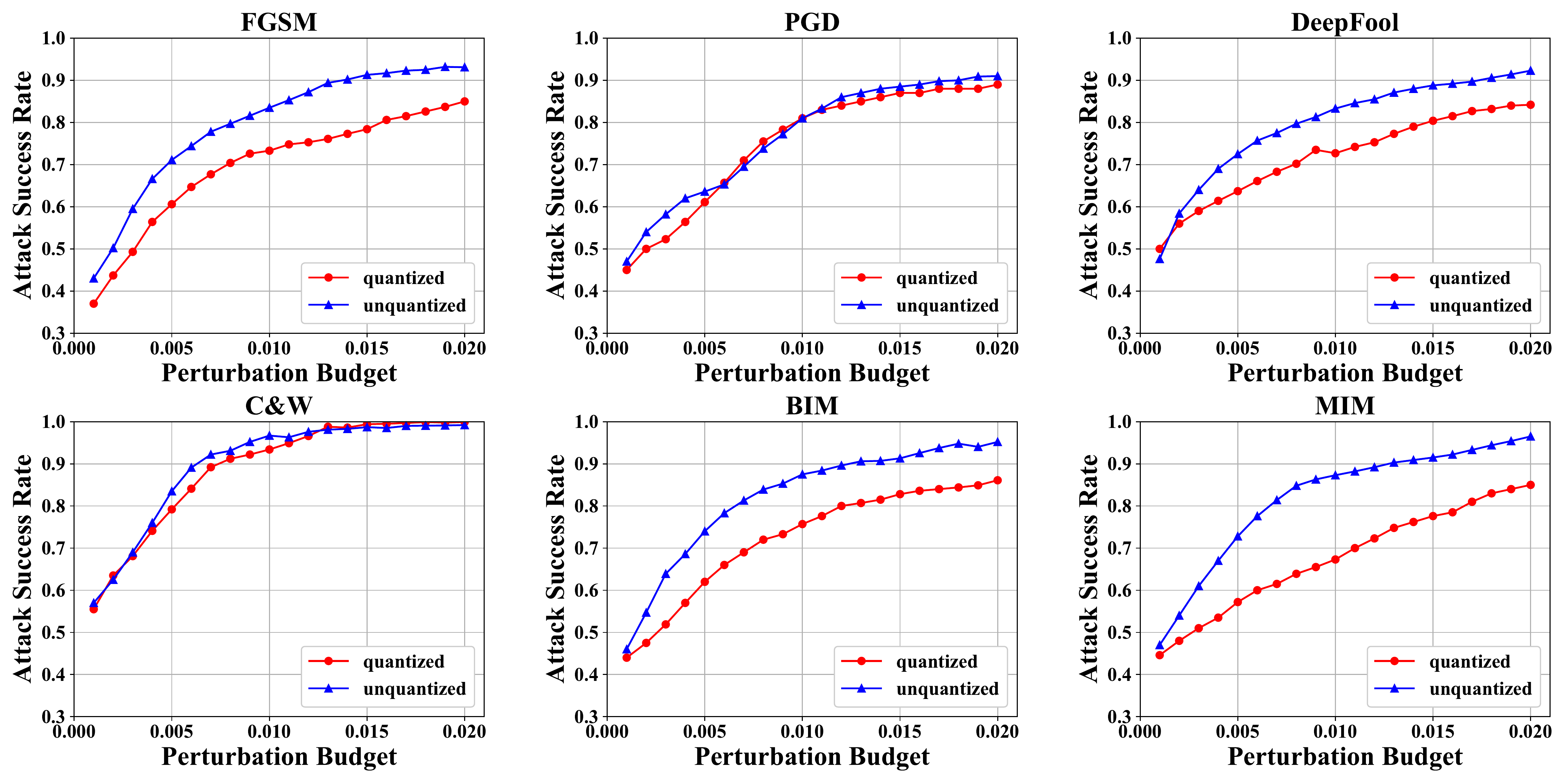}
	\caption{Comparison between quantized and unquantized models with different perturbation budgets against untargeted white-box attacks under the $l_\infty$ norm.}
	\label{fig:quan_vs_unquan}
	\vspace{-5mm}
\end{figure*}

Using these \dzz{6} attack algorithms with best-practice parameters described in RealSafe~\cite{dong2020benchmarking} and DeepSec~\cite{ling2019deepsec}, \dzz{we make a performance comparison on the same real-world model using different attack algorithms.} For instance, we choose the most-used real-world model MobileNet as our target and the results are shown in Figure~\ref{fig:diff_algo}.
We also select three popular DNNs for testing--ResNet, MnasNet, and InceptionNet.
For each model, we compute \todo{the $ASR_s$} of six attack algorithms on the same dataset from ImageNet.
We find that the most effective algorithm against real-world models is method C\&W, but it is slower compared to other algorithms.

Since \dzz{40.55\% (238/587)} of models have been quantized before being deployed on edge devices, we conduct another experiment to quantify the influence of model quantization and determine the main reason for the \rev{$ASR_m$} gap. 
\dzz{We randomly select \dzz{20} unquantized models and also convert them into \dzz{20} quantized ones to perform 6 white-box attacks. The models are all from the real world and the results are shown in Figure~\ref{fig:quan_vs_unquan}.}
We attack each model at different perturbation budgets, ranging from 0 to 0.02 with a stride of 0.001, and find that the \rev{$ASR_s$} of the quantized models is lower than that of the unquantized models. 
At the perturbation budget 0.02, the $ASR_s$ difference between the quantized model and the unquantized model can be 5-10\%.
The root cause of this problem is that quantized models have a certain degree of gradient masking~\cite{papernot2017practical}.

Moreover, quantized models have poor transferability with AEs generated from unquantized models.
\rev{
	This phenomenon is consistent with prior studies. For example, Galloway et al.~\cite{galloway2017attacking} point out that quantized networks have better robustness against adversarial attacks. Bernhard et al.~\cite{bernhard2019adversarial} state that quantized models have a quantization shift phenomenon which ruins the adversarial effect.
	Although \cite{lin2019defensive} draws a different conclusion from us, their work is mainly based on only two models (i.e., VGG-16 and Wide ResNet) with AE attack setting ($\epsilon$ ranges from 1/255 to 9/255), and it only takes into account activation quantization.
	Additionally, based on our findings, real-world quantized models mostly adopt the post-training 8-bit integer quantization method~\cite{tflite_optimize} which is parameter-quantized.} 
It is observed that the C\&W algorithm performs the best with regard to quantization, causing about \rev{1-2\% $ASR_s$} loss.
Because the C\&W method essentially turns the generation of AEs into an optimization problem, there is little dependence on the gradient information of the original model.

\emph{Summary. It is concluded that model quantization can to some extent raise the robustness of a DL model, which results in a lower $ASR_s$ in real-world models.}

\vspace {3pt}\noindent\textbf{Black-box adversarial attacks.}
\tool{} performs three types of attacks--transfer-based, score-based and decision-based attacks on black-box models.
Table~\ref{tab:exp} shows that \dzz{43} models are vulnerable to NES attack, \dzz{33} models are vulnerable to boundary attack, and \dzz{12} models to transfer attack. 
Finally, we find that \dzz{48.86\% (43/88)} of successful black-box attacks use NES, which is most effective.
The transfer attack of the black-box model has two requirements to train a substitute model, suitable dataset, and model query.
We have no limit on the number of queries. In most cases, the internal structure and parameters of the model are not clear.

\begin{figure*}
\centering
\includegraphics[width=0.75\textwidth]{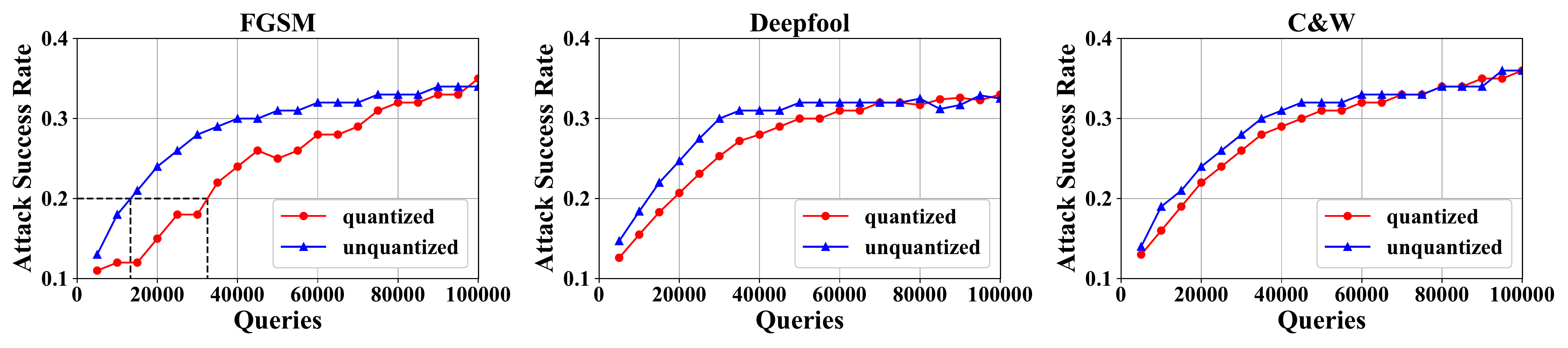}
\caption{Comparison between quantized and unquantized models with \re{the number of queries} against untargeted transfer-based attacks \re{with $\epsilon=0.06$ (16/255)} under the $l_\infty$ norm.}
\label{fig:query_times}
\vspace{-3mm}
\end{figure*}

We find that black-box quantized models need more queries to train a transfer model to reach the same \rev{$ASR_s$} on unquantized models.
As well known, it is difficult to train a substitute model, because we do not know what model architecture to train with.
With more queries, the substitute model may have more abilities to learn the original model's feature.
We choose three different white-box algorithms to attack substitute models, and the results are shown in Figure~\ref{fig:query_times}.
When the query number reaches \dzz{32,500}, the model accuracy decreases as the same level as the unquantized model with \dzz{13,333} queries, additional \dzz{19,167} queries bring extra time and resource consumption.
\re{The upper bound of $ASR_s$ reflects the consistency of the substitute model with the original. With fewer queries, the substitute model is more consistent with unquantized models so the generated AEs exhibit stronger transferability. Over the number of queries, the consistencies of the unquantized and quantized model reach a similar upper bound and thereby the gap decreases.}

\vspace {3pt}\noindent\textbf{White-box Case Study.} 
\texttt{\small ``Nsfw.pb''} is found in \dzz{19} social and video apps, e.g., an app named \texttt{\small HOLLA} with 10+ million installs. It is a not-suitable-for-work (NSFW) image classifier with two labels (non-NSFW or NSFW).
By using \tool{}, we change a dog image's label from the non-NSFW to NSFW using DeepFool with $\epsilon=0.14$ under the $l_\infty$ norm.
This kind of attack is applicable in many scenarios, like bypassing the detection of received images on victim's phone. On the contrary, if the app user acts as an attacker, it is more practical to upload the AEs of NSFW images rather than real NSFW images by intentionally disabling NSFW detection in app. It is because NSFW images can be likely blocked by either the server or the receivers' detectors. Besides, many apps have adopted app integrity check (e.g., SafetyNet~\cite{safetynet}) that makes it harder to manipulate one app nowadays~\cite{ibrahim2021safetynot}.

\vspace {3pt}\noindent\textbf{Black-box Case Study.} 
We test a black-box model which is developed with NanoNet~\cite{nanonet}.
Its usage is to detect the traffic lights and construction zones from the pictures captured by the dash-cam app, a popular app called \texttt{\small Nexar}~\cite{nexar}, with about 1 million downloads.
Its task is to identify and classify traffic lights in real-time scenarios. 
It has four output labels: empty, red, green, and yellow. 
We generate a training dataset from Internet, and train a substitute model with MobileNet after \dzz{50,000} queries. After that, we use targeted white-box method to generate \dzz{10} AEs, and find that \dzz{8} of them have the same attack effects in the original model.
This kind of AEs can bring considerably severe consequences. Supposing there are two cars on the road, the attacker can put an adversarial stop sign or red light sign on the front car. So that the rear car, the victim who is using this dash-cam app, will make wrong driving decisions.

Another example is the combination model MTCNN~\cite{zhang2016joint} for face detection.
As a well-known cascade CNN face detection system, \tool{} finds \dzz{31} apps that use the MTCNN model to recognize human face.
We choose the app named \texttt{\small TrafficPerson}, which contains three sub on-device models named \texttt{\small ``Onet.param''}, \texttt{\small ``Pnet.param''}, and \texttt{\small ``Rnet.param''} implemented on the NCNN framework. 
This framework only provides native interfaces, which is difficult to build such a test environment and attack the model outside the app. 
Therefore, \tool{} uses the black-box method to dynamically call the API provided in the app to perform model inference. \tool{} randomly generates a 20 $\times$ 20 patch image, and pastes it on the face area of original image.
Since there is no limit of query numbers, \tool{} perturbs the patch pixels for \dzz{5,000} epochs. After that, we find \dzz{5} patched images can hide the original face and be used in a physical attack to bypass the face detection.
In the physical world, it is much cost-effective to add a patch on the face to fool the detection system and the method by adding perturbations is imperceptible by human, and thus more severe.

\section{Suggestions for Developers}
\label{sec:guide}

On-device models are important intellectual properties to app developers and model providers.
On one hand, improper protection of on-device models may attract attackers to launch attacks against legitimate app function, leading to various consequences.
On the other hand, the misuse of such on-device models may cause damage to the revenue of the model providers, as attackers can plagiarize the model through infinite queries and building a substitute model.
However, it is well recognized that the arms race between attack and defense never ends. 
It is appealing for defenders to raise the barrier for attackers so that the benefit from the attack cannot offset the attack cost. 
\re{Here we provide three suggestions against model stealing, misuse and AE attacks.}

\vspace {3pt}\noindent\textbf{1. Avoid clear text on-device models.}
Our analysis tool relies on a lot of useful information provided by the clear text on-device models.
Therefore, to avoid being analyzed, the straightforward way is to well protect on-device model files.
Obfuscation is one cost-effective approach by obscuring any meaningful text in the stored model files. 
\re{Encryption can even make the model exist only in memory, and users cannot identify the model by looking at the storage.
We found that \re{70} apps take the defense. In these apps, we can only extract models from 18.57\% (13/70) of them. In contrast, for unprotected apps, 97.80\% (241/246) of them can be successfully analyzed for model extraction, demonstrating the defense's effectiveness. We further analyzed the failed apps and found that 23 apps use obfuscation to make plaintext unrecognizable. While we cannot directly use string matching to locate the code for model loading, we can still use dynamic analysis to find the code. However, for the rest of the 34 apps using encryption, it is really hard for us to extract the models. So heavy-weight encryption/obfuscation can make the analysis very difficult. A developer can even split the model into two parts and encrypt them~\cite{florian2019slalom}. \re{When using the model, users dynamically decrypt the two parts and combine them to form a whole DL model, which can increase the difficulty of model extraction.}
Obfuscation and encryption are two common techniques for protecting apps. Developers only need to follow standard methods to deploy the two techniques.}

\vspace {3pt}\noindent\textbf{2. Prevent the misuse of on-device models.}
On-device models are valuable assets, and hence the approaches of protecting assets can be applied directly, e.g., authentication.
Authentication ensures that only authenticated users/apps are permitted to use the on-device models.
This requires the collaboration between app developers and model providers.
For app developers, this can be performed by requiring login and user token before the use of their on-device models.
For model providers, secure license distribution from server side and proper license management from client side are necessary before the use of their models.
\re{
According to our evaluation, \re{47} apps take advantage of this defense, and they are all Type-C. Recall that the Type-C model uses closed source frameworks. That is to say, even if we successfully extract the models, we cannot directly load them for white-box analysis. So the only choice is to treat the apps as black-box and query the models using various input/output pairs. However, for \re{35} apps that use the defense, we cannot obtain proper licenses or tokens and cannot execute queries.
For other Type-C models without the defense, we can successfully conduct black-box attacks.}
For example, the AppLock app with about 1 million downloads, uses Sensory's face liveness detection model. 
This model lacks a license manager to verify user access to the model. 
An attacker can just copy the model and the library to steal this closed source model and rebuild another face detection app. 
\re{One challenge of deploying the defense is securing licenses well. Once the license is compromised, malicious users can query the model and perform black-box attacks. Therefore, it is recommended to store licenses on the cloud and update authentication credentials regularly.}
Besides, recent studies use TEE to protect mobile models from AE attacks by concealing model loading and execution (e.g., OMG~\cite{bayerl2020offline}).

\vspace {3pt}\noindent\textbf{3. Build more robust on-device models.}
Last but not the least, it is important for both app developers and model providers to bear in mind the necessity of enhancing the robustness of the on-device model itself.
It is shown that the robustness can be achieved using various techniques, such as adversarial training~\cite{madry2018towards, zhang2019theoretically} and input transformation~\cite{guo2018countering}. 
There are also approaches~\cite{ma2018characterizing, Feinman2017DetectingAS} to detect whether the input is an AE.
\re{To evaluate the efficacy of these three defenses, we randomly select \re{20} real-world models and compare the attack success rates ($ASR_s$) before and after the defenses. Specifically, we employ PGD-AT~\cite{madry2018towards} and TRADES~\cite{zhang2019theoretically} for adversarial training. The $ASR_s$ drops by 37.5\% on average, indicating that 37.5\% of adversarial examples cease to be effective. As for input transformation, we use JPEG compression~\cite{guo2018countering} and Pixel Deflection (PD)~\cite{Prakash2018DeflectingAA}, which reduces the $ASR_s$ by 17.0\% on average. For AE detection, we take two popular methods--Local Intrinsic Dimensionality (LID)~\cite{ma2018characterizing} and Kernel Density Estimation (KDE)~\cite{Feinman2017DetectingAS} with the default settings. The average detection ROC-AUCs are 82.1\% and 71.5\%, respectively. Detailed experiment settings and results are shown in Appendix D
.}

\re{We also compare the time cost of the three defenses. For adversarial training, the average training time is 2.8 hours. Input transformation does not require additional time to retrain the models. Its cost is mainly from dynamic execution, which on average takes about \re{0.03} seconds to process an image.
AE detection usually requires to extract features of AEs \re{and} models and train a detector for recognizing abnormal input, which takes us 1.2 hours on average. The extra time comes from detectors' inference of the given input, which averagely takes about \re{0.5} seconds.}

\re{The main challenge of deploying the defenses is to minimize the impact on normal inputs. Adversarial training requires updating the model, which may affect the model's accuracy; input transformation requires changing the inputs, which may make the original input unrecognizable; AE detection can also classify normal inputs as adversarial examples. Therefore, developers should deploy defenses carefully so as not to affect the normal requests of users. 
}

\re{\emph{Summary. We find that the first two suggestions have already been deployed in some apps to prevent attackers from extracting and misusing models.
Although there are no evidences from our study that the contemporary apps have leveraged measures to prevent adversarial examples, the investigated defenses including adversarial training, input transformation and AE detection have shown their considerable potential in model enhancement. Besides, the additional time overhead is less than half a second for one inference.}}

%\vspace {7pt}
\section{Discussion}
\label{sec:discuss}

\re{\noindent\textbf{Real-world vs. Academic models.}}
\re{After extracting and evaluating these on-device models, we make a comparison between real-world models and conventional models.}
\re{Real-world models exhibit significant differences from those in academia based on our study. Specifically, real-world models are prone to employing:
\X1 cost-efficient network architectures like MobileNets, EfficientNet, which negatively impact prediction accuracy (see Figure~\ref{fig:model_accuracy}). In the RWM dataset, these models' average file size is 4.35 MB.
\re{In particular, there are 84 MobileNets, 18 SqueezeNets, 13 EfficientNets in the RWM dataset.}
\X2 layers obfuscation or custom operators designed for edge devices and hardware accelerators (e.g., TPU, NPU) in Section~\ref{sec:system-threat}. There are 12 kinds of custom operators found in the RWM dataset, e.g., MaxPoolingWithArgmax2D, MaxUnpooling2D, StrSim, and TextEncoder3.
\X3 model quantization (40.55\% as stated in Section~\ref{sec:ae-threat}), model compression and pruning for optimization.
Such differences could inspire interesting research questions.}

\noindent\re{\textbf{Generalizability Analysis.} For the on-device models in the RWM dataset, we conduct an empirical study on model characteristics and protections, and a quantitative analysis of threats from adversarial examples. In particular, we unveil in the overall view model inputs, tasks and optimizations. We also identify five system-level protections and three strategies to enhance models against physical theft. All the above is irrelevant to model types, so the findings can be generalized to other scenarios. Although we only take into account the image-based models in RQ3, this type of models takes up over 70\%, so that the conclusions in this part can well depict the current threats in DL apps.}

\noindent\textbf{On-device training.}
\rev{As of TensorFlow 2.7 (released in Nov, 2021), it supports on-device training~\cite{ondevicetraining}, which enables one model to be updated with trainable weights from the device or end user. From our dataset we find no model existing this support yet, and it will bring great help for attackers to perform white-box attacks. On the contrary, it also offers a possibility for models to raise their robustness by retraining on device at any time.}

\noindent\textbf{Limitations.}
During the model extraction in Section~\ref{sec:method-extract}, there may be some false negatives due to the model protection adopted by app developers. 
For example, for the app that adopts obfuscation techniques, \tool{} based on semantic search may not be able to find any model-related strings in the app's bytecode or lib export functions. 
Moreover, the file formats of protected models in some apps are with suffixes like \textit{.bin} or \textit{.data} instead of common model file format like \textit{.tflite}. 

\noindent\textbf{Future work.}
In this work, we only try several mainstream attacks which are experimentally effective. 
There are more and more powerful attacks being proposed.
To keep up with the arm race, we will experiment with more new attack methods in the future, e.g., recently proposed optimized boundary attack~\cite{brendel2017decision}. 
It is an interesting extension to our work to study models with more diverse functions, such as audio recognition and natural language processing, other than image-related functions.
Even so, we believe \tool{}, which facilities model extraction, interface reasoning and automated testing, can benefit the security assessment of mobile DL models.

%\vspace {7pt}
\section{Related Work}
\label{sec:related}

\begin{comment}
\vspace {3pt}\noindent\textbf{DL model in mobile applications.}
In recent years, DL has demonstrated its superior performance in many areas~\cite{krizhevsky2012imagenet, hinton2012deep, andor2016globally, cai2016unified}.
More and more mobile applications rely on DL to complete complicated tasks in an intelligent manner. 
There have been efforts to improve the performance of on-device models using model compression~\cite{liu2018demand, wu2016quantized} and hardware customization~\cite{chen2014diannao, han2016eie}. 
As the on-device model inference has been practically adopted on mobile devices, it is of great significance to analyze and understand the potential security threat in these on-device models~\cite{xu2020survey}.
Ignatov et al.~\cite{ignatov2019ai, ignatov2018ai} evaluate different mobile hardware accelerators for on-device model inference and present the first real-world benchmark performance of different mobile SoCs.
Xu et al.~\cite{xu19first} give a glance at how DL techniques are deployed in smartphone apps.
Their work uses statistic methods to provide insightful results, for example, how many top apps use DL techniques, what do the apps use DL for, what is the average size of on-device models, are the apps using any optimization techniques, etc.
Based on the revealed results, they provide implications for multiple stakeholders of the mobile DL ecosystem.
Different from their work, our study is to understand the real-world adversarial attacks on ML models.
\end{comment}

\noindent\textbf{On-device model evaluation.}
As the on-device inference has been practically adopted on mobile devices, it is of great significance to analyze and understand the potential security threats in these on-device models~\cite{xu2020survey}.
Ignatov et al.~\cite{ignatov2019ai, ignatov2018ai} evaluate different mobile hardware accelerators for on-device model inference and present the first real-world benchmark performance of different mobile SoCs.
Xu et al.~\cite{xu19first} give a glance at how DL techniques are deployed in smartphone apps.
Their work uses statistical methods to provide insightful results, for example, how many top apps use DL techniques, what these apps use DL for, what is the average size of on-device models, and whether the apps are using any optimization techniques, etc.
Based on the revealed results, they provide implications for multiple stakeholders of the mobile DL ecosystem.
However, it is worth noting that though plenty of toolboxes, such cleverhans~\cite{papernot2018cleverhans} and ART~\cite{art2018}, are developed for practically evaluating DL models using traditional DL framework, few of work has been done to evaluate on-device models that are built upon mobile DL frameworks. 
\emph{Different from their work, our study is to understand the gap between real-world adversarial attacks and academic attacks on DL models, and show robustness and threats in real-world models.}

\noindent\textbf{Attacks on DL apps.}
Deep learning has been witnessed to be suffering from many attacks, for example adversarial attacks~\cite{goodfellow2014explaining, papernot2016limitations, moosavi2016deepfool, carlini2017towards, chen2017zoo, papernot2017practical}.
Attacks aim to find AEs to force a machine learning system to produce erroneous outputs. 
This type of attacks can be done in both white-box manner~\cite{goodfellow2014explaining, papernot2016limitations, moosavi2016deepfool, carlini2017towards} and black-box manner~\cite{chen2017zoo, papernot2017practical}. 
More specifically, white-box attacks can be categorized into gradient-based methods~\cite{goodfellow2014explaining, papernot2016limitations} and optimization-based methods~\cite{moosavi2016deepfool, carlini2017towards}.
Due to the lack of sufficient model information, black-box attacks are achieved using scored-based methods to estimate the gradient~\cite{chen2017zoo}, or using the transferability feature of the adversarial input between an accessible model and the victim black-box model~\cite{papernot2017practical}.
In addition, there are attacks against specific critical DL apps, such as biometric authentication system~\cite{Chen2019BiometricAU}, liveness detection system~\cite{zhao2020resilience} and malware detection~\cite{pierazzi2019intriguing}.

%\vspace {7pt}
\section{Conclusion and Outlook}
\label{sec:conclusion}

This work is the first measurement work looking at security risks of learning models deployed in mobile apps. We have found that \dzz{5,573 (8.90\%)} of \dzz{62,583} apps from the market have been equipped with DL technology to enhance their performance, while \dzz{568 (10.19\%)} of them have local on-device models inside the APK files. Many on-device models are unprotected and vulnerable to adversarial attacks. We proposed a novel method to get model interface of DL models and build model testing environment to generate the AEs.
Benefiting from our model analysis, \dzz{47.35\%} of the models can be successfully attacked by current AE generation algorithms. 
We also found that real-world models with quantization are more robust than academic ones. In the end, we provided insightful suggestions on deploying DL models and defending AE attacks for developers.

\section*{acknowledgements}
We thank all the anonymous reviewers for their constructive feedback.
IIE authors are supported in part by the National Key R\&D Program of China (2020AAA0140001), NSFC (U1836211, 61902395), Beijing Natural Science Foundation (No.M22004), the Anhui Department of Science and Technology under Grant 202103a05020009,  Youth Innovation Promotion Association CAS, Beijing Academy of Artificial Intelligence (BAAI) and a research grant from Huawei.
%\newpage
%%
%% The next two lines define the bibliography style to be used, and
%% the bibliography file.
%\nocite{*}
\bibliographystyle{ACM-Reference-Format}
\balance
\bibliography{paper}

%%
%% If your work has an appendix, this is the place to put it.
\section*{Appendices}
\label{sec:appendix}
\renewcommand{\thesubsection}{\Alph{subsection}}

\begin{table*}[ht]

	\footnotesize
	\centering
	\caption{The regular expression list for on-device model recognition} %\guozhu{why caffe and caffe2 are duplicated, change the table }
	\label{tab:formatlist}
	
	\vspace{-3mm}
	\begin{threeparttable}
		\begin{tabular}{cccccc}
			\toprule
			
			\multirow{2}{*}{\textbf{Framework}} & \multirow{2}{*}{\textbf{Owner}} & \multicolumn{2}{c}{\textbf{File Feature}} & \multicolumn{2}{c}{\textbf{Code Feature}} \\ \cline{3-6}
			& & \textbf{Model Suffix} & \textbf{Model Format} & \textbf{Lib Name}  &  \textbf{Lib Func Name} \\
			\midrule
			Tensorflow(Lite) & Google   & .tflite\$$\vert$.lite\$$\vert$.pb\$$\vert$.pbtxt\$$\vert$ckpt\$       & TFL3    & libtensorflow$\vert$tensorflow           & - \\
			Caffe            & Caffe    & .caffemodel\$       & -       & libcaffe$\vert$caffe                       & - \\
			Caffe2           & Facebook & .pb\$$\vert$.pbtxt\$$\vert$ckpt\$ & -       & libxplat\_caffe2$\vert$xplat\_caffe2 & N\textbackslash d+caffe\textbackslash d+NetDefE \\
%			Opencv           & Opencv   & -                   & -       & libopencv$\vert$opencv                   & - \\
			MindSpore        & Huawei   & .ms\$               & -       & libmindspore-lite$\vert$mindspore        & - \\
			Paddle(Lite)     & Baidu    & .nb\$                   & -       & libpaddle$\vert$paddle                   & - \\
			MACE             & Xiaomi   & -                   & -       & libmace$\vert$mace                       & - \\
			%support models of other framworks
			Parrots          & SenseTime & -                  & -       & libst\_mobile$\vert$st\_mobile           & - \\
			XNN              & Alibaba  & .xnntflite\$        & -       & libxnn$\vert$xnn                         & - \\
			MNN              & Alibaba  & .mnn\$                   & -       & libmnn$\vert$mnn                                       & MNNNet \\
			TNN              & Tencent  & .tnnmodel\$$\vert$.tnnproto\$                  & 0FABC0002h & libtnn\_wrapper                                     & - \\
			NCNN             & Tencent  & .param\$$\vert$.cfg.ncnn\$$\vert$.weights.ncnn\$ & 7767517 & libncnn           & - \\
			AlphaFace        & DiDiTech & .bin.alg\textbackslash d\$  & -    & libalphaface$\vert$alphaface                                   & - \\
			MxNet            & Apache   & -                   & -       & libmxnet\_predict$\vert$mxnet & N\textbackslash d+mxnet\textbackslash d+EngineE \\
			%.json,.params
			Sensory          & Sensory  & .model\$                   & -       & libsmma$\vert$smma                       & - \\
			Megvii           & Face++   & -                   & -       & liblivenessdetect   & - \\
			Cognitive        & Microsoft & -                  & -       & libofflinetranslator$\vert$offlinetranslator & - \\
			% mostly use as a service?
			ONNX        & ONNX & .ort\$$\vert$.onnx\$              & -       & libonnxruntime & - \\
			\bottomrule
		\end{tabular}
	\end{threeparttable}
\end{table*}

\begin{table*}%[ht]
\renewcommand\arraystretch{1.1}
\footnotesize
\centering 
\caption{The common API list with signature for on-device model from DL frameowrks' SDK/NDK documents}
\label{tab:apilist}

\vspace{-3mm}
\begin{threeparttable}
\begin{tabular}{cccc}
\toprule
\textbf{Framework} & \textbf{Loader API} & \textbf{Inference API} & \textbf{Location}  \\
\midrule

\multirow{2}{*}{TFLite}    & Void;Interpreter(MappedByteBuffer)  & Tensor[];NativeInterpreterWrapper.run(Object[])   & \multirow{2}{*}{dex} \\
          & MappedByteBuffer;FileUtil.loadMappedFile(Object, String) & void;runForMultipleInputsOutputs(Object[], Map$\langle$Integer, Object$\rangle$) &   \\
\cline{1-4}
PyTorch   & Void;Module.load(String [assetFilePath])      & Tensor;Module.forward(IValue[]) & dex \\
\cline{1-4}
\multirow{2}{*}{NCNN}   & int;load\_param(const DataReader \&)   & \multirow{2}{*}{int;Extractor.extract(const char *, Mat \&, int)}  & \multirow{2}{*}{native C++} \\
      & int;load\_model(const DataReader \&) & & \\
\cline{1-4}
MNN       & Session;createSession(Config)                 & void;runSession(Session) & native C++ \\
\cline{1-4}
MindSpore & Boolean;Model.loadmodel(Context,String)       & Boolean;Session.runGraph(void) & dex \\
\cline{1-4}
PaddleLite & void;MobileConfig.setModelFromFile(String)   & void;createPaddlePredictor.run(void) & dex \\
\cline{1-4}
MxNet & long;creatPredictor(byte[],byte[],int,int,String[],int[][])   & void;nativeForward(long,String,float[]) & native C++ \\
\cline{1-4}
Caffe & boolean;loadModel(String,String)   & float[];predict(byte[],int,float[]) & native C++ \\
\cline{1-4}
Caffe2 & void;initCaffe2(AssetManager)   & String;classificationFromCaffe2(int,int,byte[],byte[],byte[],int,int,boolean) & native C++ \\
\bottomrule
\end{tabular}
\end{threeparttable}
\end{table*}

\subsection{DL Framework Characters}

Table~\ref{tab:formatlist} shows top 17 DL frameworks' characters, including file features and code features. \re{If an app has any features of DL framework and MIS found in the app code, we infer that the app has on-device models.}

\subsection{Inference Rules for Model Interface}
Table~\ref{tab:rules} shows reasoning rules used by \tool{} for model interface analysis. Based on the framework documents, we have collected all the related APIs and comments, and then extracted representative keywords from them. To ease the inference, we compile a list of regular expressions to recognize the information including model details, input and output.

\begin{table*}
\footnotesize
\renewcommand\arraystretch{1.2}
\centering
%\captionsetup{labelformat=empty}
\caption{Reasoning rules for model interface}
\label{tab:rules}
\vspace{-3mm}
\begin{tabular}{l|l|l} 
\Xhline{2\arrayrulewidth}
\textbf{Part}             & \textbf{Type}          & \textbf{Representative Keywords \& Regular expressions}                                                                             \\ 
\cline{1-3}
\multirow{4}{*}{ Model  } & model loading           & LOADER\_API\_LIST \& INFERENCE\_API\_LIST (See Table~\ref{tab:apilist})                                                                                                                                                                                                                                                                                                                                                                                                                                                                                                                                                                                                                                                                                                    \\ 
\cline{2-3}
                          & model task              & \begin{tabular}[c]{@{}l@{}}TASK\_DICT = \{`classification' ~ ~ ~:~[`classif*', `softmax', `recog*', ...], `object~detection' :~[`ssd', `onet.', `rnet', `pnet', `detect*', `bound*', ...],~\\~ ~ ~ ~ ~ ~ ~ ~ ~ ~ ~ ~ ~ ~ ~ ~ ~`pose~detection' ~ :~[`pose', ...], ~ `segmentation' ~ :~[`segment', `outline', ...] `stylize' ~ ~ ~ :~[`styl*', `gan', ...], `sequence~predict' :~[`rnn', `lstm', ...]~\},~\\TASK\_NAME = [`(\textbackslash{}w+)\{NOUN\_LIST\}', ...], \\NOUN\_LIST ~= [`detector', `detection', `classifier', `classification', `inference', `predictor',~`prediction', `recognizer', `recognition', ...]\end{tabular}   \\ 
\cline{2-3}
                          & model optimization      & OPT\_LIST ~ ~ = [`optimiz*', `quant*', ...]                                                                                                                                                                                                                                                                                                                                                                                                                                                                                                                                                                                                                                                                                     \\ 
\cline{2-3}
                          & model arch/backbone     & ARCH\_LIST ~= [`inceptionresnet', `resnet', `mobilenet', `vgg', `mnas', `squeeze', `efficient', ...]                                                                                                                                                                                                                                                                                                                                                                                                                                                                                                                                                                                                            \\ 
\cline{1-3}
\multirow{3}{*}{Input}    & input format            & \begin{tabular}[c]{@{}l@{}}INPUT\_SOURCE\_API\_LIST = [`createBitmap', `createScaledBitmap', ...]\\ INPUT\_DICT = \{`image'~:~[`img', `image', `camera', `picture', `open camera \textbar{} video \textbar{} album', ...],\\~ ~ ~ ~ ~ ~ ~ ~ ~ ~ ~ ~ ~ `audio'~:~[`audio', `speech', `voice', `open microphone', ...], `text'~~:~[`translat', `nlp', ...]\}\end{tabular}                                                                                                                                                                                                                                                                                                                                                                                    \\ 
\cline{2-3}
                          & input preproc method & PREPROC\_API\_LIST = [`resize', `norm*', `rescale', `preprocess', `prepare', ...]                                                                                                                                                                                                                                                                                                                                                                                                                                       \\ 
\cline{2-3}
                          & input preproc param  & PARAM\_LIST = [`mean', `std' , ` \textgreater{}\textgreater{} \textbackslash{}d+)~\&~255)', `with (\textbackslash{}d+) (\textbackslash{}d+)', ...]                                                                                                                                                                                                                                                                                                                                                                                                                                                                                                                                                                                       \\ 
\cline{1-3}
\multirow{2}{*}{Output~}  & output handler  &   HANDLER\_API\_LIST = [`argmax', `mapping',  `render', `setuiview*', ...]                                                                                                                                                                                                                                                                                                                                                                                                                                                                                                                                                                                                                                                                               \\ 
\cline{2-3}
                          & output label            & \begin{tabular}[c]{@{}l@{}}OUTPUT\_LIST = [`enum', `switch', `case', `result', `score', `output', `label', `confidence', ...]\\CODE\_LIST ~ ~ = [`enum class (\textbackslash{}w+)', `switch(.*) case \textbackslash{}w+*', ...]\end{tabular}                                                                                                                                                                                                                                                                                                                                                                                                                                                                           \\
\Xhline{2\arrayrulewidth}
\end{tabular}
\end{table*}

\subsection{Attack Validation}
\label{sec:attack-validate}

Ideally, the validation of AEs can be accomplished by installing and running the target app, passing AEs to the apps and obtaining the inference results. 
However, it is non-trivial to make the model receive AEs since the MIS may be difficult to reach. To address this, \tool{} employs DroidBot~\cite{droidbot} to traverse code paths and invoke the model. 
On the other hand, it may retain unsuccessful to pass AEs to the model. As a result, we propose two strategies to overcome this difficulty.

\begin{lstlisting}[language=java,frame=trBL,caption=The instrument code when testing AEs .,abovecaptionskip=10pt,belowcaptionskip=0pt,captionpos=b, label={lst:instrument_code},
  keywordstyle=\bfseries\color{green!40!black},
  commentstyle=\itshape\color{purple!40!black},
  identifierstyle=\color{blue},
  stringstyle=\color{orange},
  basicstyle=\footnotesize\ttfamily,
  breaklines=true,
  numbers=left,
  numbersep=-5pt,
  stepnumber=1,]
  invoke-static {}, Landroid/os/Environment;->getExternalStorageState()Ljava/lang/String;
  move-result-object v0
  ...
  const-string v1, "adversary.png"
  invoke-virtual {v0, v1}, Ljava/lang/StringBuilder;->append(Ljava/lang/String;)Ljava/lang/StringBuilder;
  ...
  invoke-static {v0}, Landroid/graphics/BitmapFactory;->decodeFile(Ljava/lang/String;)Landroid/graphics/Bitmap;
\end{lstlisting}

\begin{lstlisting}[language=java,frame=trBL,caption=The normalization code of nanoNet.tflite (in Java), abovecaptionskip=10pt,belowcaptionskip=0pt,captionpos=b, label={lst:precode},
keywordstyle=\bfseries\color{green!40!black},
commentstyle=\itshape\color{purple!40!black},
identifierstyle=\color{blue},
stringstyle=\color{orange},
basicstyle=\footnotesize\ttfamily,
breaklines=true,
numbers=left,
numbersep=-5pt,
stepnumber=1,]
  private static final float IMAGE_MEAN = 127.5f;
  private static final float IMAGE_STD = 128.5f;
  public void addPixelValue(int i) {
   this.imgData.putFloat((((float)((i >> 16) & 255)) - IMAGE_MEAN) / IMAGE_STD);
   this.imgData.putFloat((((float)((i >> 8) & 255)) - IMAGE_MEAN) / IMAGE_STD);
   this.imgData.putFloat((((float)(i & 255)) - IMAGE_MEAN) / IMAGE_STD); 
  }
\end{lstlisting}

\begin{itemize}[leftmargin=*]
	\item \textbf{Dynamical parameters modification.} 
	Dynamically modifying the input parameters of the model with AEs, then observing the inference results at runtime. 
	Considering the condition of whether performing AEs on the app is consistent with the effect of external AE attacks. 
	If they are consistent, it means that the app is indeed vulnerable to AE attacks. 
	Otherwise, it means that the app has an additional detection mechanism, or the behavior of the same model in different environments is inconsistent.
	\item \textbf{App repackaging.} 
	%In most attack scenarios, the attacker cannot have such a powerful ability to repackage the app.
	Due to our assumptions in Section~\ref{sec:method}, repackaging is possible for attackers to verify AE attacks locally by: \X1 putting AEs into the shared storage on the mobile phone; \X2 decomposing the app at the location for DL function invocation; \X3 adding the code (see List~\ref{lst:instrument_code}) and permissions to replace the original model input by the AE in the external storage; \X4 after repacking and resigning the app, loading the app and observing the behaviors of the model used in the app. 
\end{itemize}

\begin{table*}[ht]
\renewcommand\arraystretch{1.15}
\footnotesize
\centering % full qualified name
\caption{The issues found from real-world apps with \tool{}}
\label{tab:buglist}

\vspace{-3mm}
\begin{threeparttable}
\resizebox{\textwidth}{!}{
\begin{tabular}{cp{2.5cm}lp{7cm}ccc}
\toprule
\textbf{App} & \textbf{Issue} & \textbf{Model Task} & \textbf{Description} & \textbf{Reported} & \textbf{Confirmed} & \textbf{Fixed} \\
\midrule

App1 & Cipher key in plaintext & Stylization & The key to decrypt model is stored as plaintext in the dex code. & Y & Y & N \\
\cline{1-7}
App2 & Cipher key in plaintext & Receipt Recognition & The key to decrypt model is stored as plaintext in the native code. & Y & - & N \\
\cline{1-7}
App3\tnote{1} & Local authentication token & Face Detection & Before using the model, the app locally authenticates with a token, but the token is easily to forge. & Y & N  & Y \\
\cline{1-7}
App4 & Model license exposure & Face Recognition & The model license file is exposed which lets anyone be able to utilize the on-device model. & Y & - & Y\\
\cline{1-7}
App5 & Redundant models & Classification & There are two identical models stored in the app, one is encrypted but the other is plaintext.  & Y & - & N \\
\cline{1-7}
App6 & Unsafe dynamic model loading & Recommendation & The model is downloaded at runtime from a remote link coded in app. So anyone can obtain models without authorization. & Y & - & Y\\
\cline{1-7}
App7 & Interface exposure & Road Condition Recognition & The road condition information recognized by the DL model is regularly uploaded to the public web page (sometimes accompanied by photos), revealing sensitive information of the users. & Y & - & Y \\

\bottomrule
\end{tabular}}
\begin{tablenotes}
\item [1] We find that an older version of this app has an authentication bypass during model use. However, in its latest version, the authentication process has been hardened with code \\that fixes the vulnerability.
\end{tablenotes}
\end{threeparttable}
\end{table*}

\subsection{Evaluation of AE Defenses}
\label{sec:defense-eval}

\re{We deploy three defenses including adversarial training, input transformation and AE detection on \re{20} randomly selected on-device models.
For each model, we perform three attacks including FGSM, DeepFool and C\&W with $\epsilon=0.03$ under the $l_\infty$ norm.
For each defense strategy, we choose two methods that are publicly available and well evaluated by~\cite{ling2019deepsec, dong2020benchmarking}.
For each method, we use the best-performing parameters from their papers:}
\begin{itemize}[leftmargin=*]
	\item \vspace {2pt}\re{\textbf{Adversarial training.} We adopt PGD-AT~\cite{madry2018towards} and TRADES~\cite{zhang2019theoretically}.
	For PGD-AT, we use 5 iterations to generate AEs for a robust training.
	For TRADES, we use 10 iterations to add perturbations.}
	
    \item \vspace {2pt}\re{\textbf{Input transformation.} We adopt Image Transformations (IT)~\cite{guo2018countering} and Pixel Deflection (PD)~\cite{Prakash2018DeflectingAA}.
    For IT, we use JPEG compression to reduce adversarial perturbations in the inputs.
    For PD, we set window size 
     = 10, sigma = 0.04, and use wavelet denosier as default.}
    
    \item \vspace {2pt}\re{\textbf{AE detection.} We adopt Local Intrinsic Dimensionality (LID)~\cite{ma2018characterizing} and Kernel Density Estimation (KDE)~\cite{Feinman2017DetectingAS}.
    For KDE, we use the features of the last hidden layer to calculate the kernel density estimate. We set bandwidth = 0.1 and Gaussian noise standard deviation 0.388 for FGSM, 0.178 for DeepFool, 0.02 for C\&W.
    For LID, we set the number of the nearest neighbours k = 10 to extract characteristics.
    The difference between these methods is mainly attributed to the extracted features. After getting the features, we use Logistic Regression to train a binary classifier for adversarial examples detection as in the papers.} 
\end{itemize}

\begin{table}[H]
\footnotesize
\centering
\caption{Evaluation of three AE defenses on 20 on-device models with $\epsilon=0.03$ under the $l_\infty$ norm}
\label{tab:ae-defense}
\begin{threeparttable}
\begin{tabular}{c|c|cc|cc|cc}
\Xhline{2\arrayrulewidth}
\multirow{2}{*}{\textbf{Attacks}}                     & \multirow{2}{*}{\textbf{Orig.}} & \multicolumn{2}{c|}{\textbf{Adv. Training}}                          & \multicolumn{2}{c|}{\textbf{Transformation}}                          & \multicolumn{2}{c}{\textbf{AE Detection}}                          \\ \cline{3-8} 
                                      &                           & \multicolumn{1}{c|}{PDG-AT} & TRADES                  & \multicolumn{1}{c|}{IT} & PD                  & \multicolumn{1}{c|}{LID} & KDE                 \\ \Xhline{2\arrayrulewidth}
FGSM                                  &       86.8\%                    & \multicolumn{1}{c|}{53.7\%}    &     45.6\%                  & \multicolumn{1}{c|}{81.2\%}    &       63.6\%                & \multicolumn{1}{c|}{84.3\%}    &    73.8\%                 \\ %\hline
DeepFool                              &     92.9\%                  & \multicolumn{1}{c|}{56.6\%}    &      47.6\%                 & \multicolumn{1}{c|}{85.2\%}    &     69.3\%                  & \multicolumn{1}{c|}{80.4\%}    &      68.6\%                \\ %\hline
C\&W                                  &     99.2\%                     & \multicolumn{1}{c|}{71.2\%}    &      57.9\%                 & \multicolumn{1}{c|}{84.7\%}    &   71.6\%                    & \multicolumn{1}{c|}{81.8\%}    &    72.1\%                  \\ \hline
\multicolumn{1}{c|}{\textbf{Average}} & \multicolumn{1}{c|}{93.0\%}     & \multicolumn{1}{c|}{60.5\%}    & \multicolumn{1}{c|}{50.4\%} & \multicolumn{1}{c|}{83.8\%}    & \multicolumn{1}{c|}{68.2\%} & \multicolumn{1}{c|}{82.1\%}    & \multicolumn{1}{c}{71.5\%} \\ \Xhline{2\arrayrulewidth}
\end{tabular}%}
\begin{tablenotes}
\item [1] Column ``Orig.'' shows the average $ASR_s$ per attack on the original models. For adversarial training and input transformations, we present the average $ASR_s$ after applying these two defenses. As for AE detection, we show the average ROC-AUC of these two algorithms--LID and KDE.
\end{tablenotes}
\end{threeparttable}
\end{table}

\begin{figure}[ht]
	\centering
\includegraphics[width=0.3\textwidth]{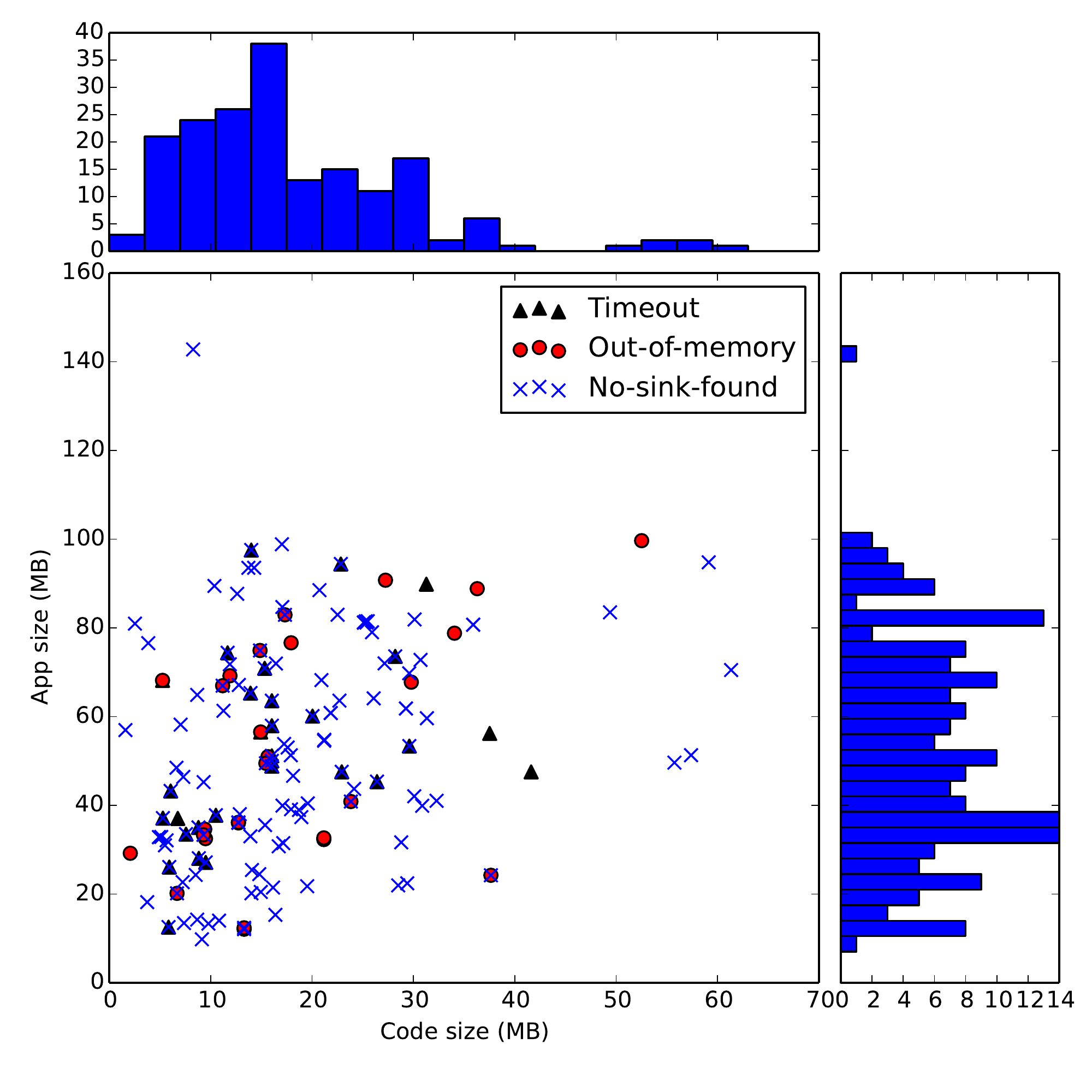}
	\caption{Statistics of the file size (app size and DEX code size) of DL apps that cannot be run by Flowdroid.}
	\label{fig:failed_app}
	%\vspace{-5mm}
\end{figure}

\begin{figure}[H]
	\centering
\includegraphics[width=0.3\textwidth]{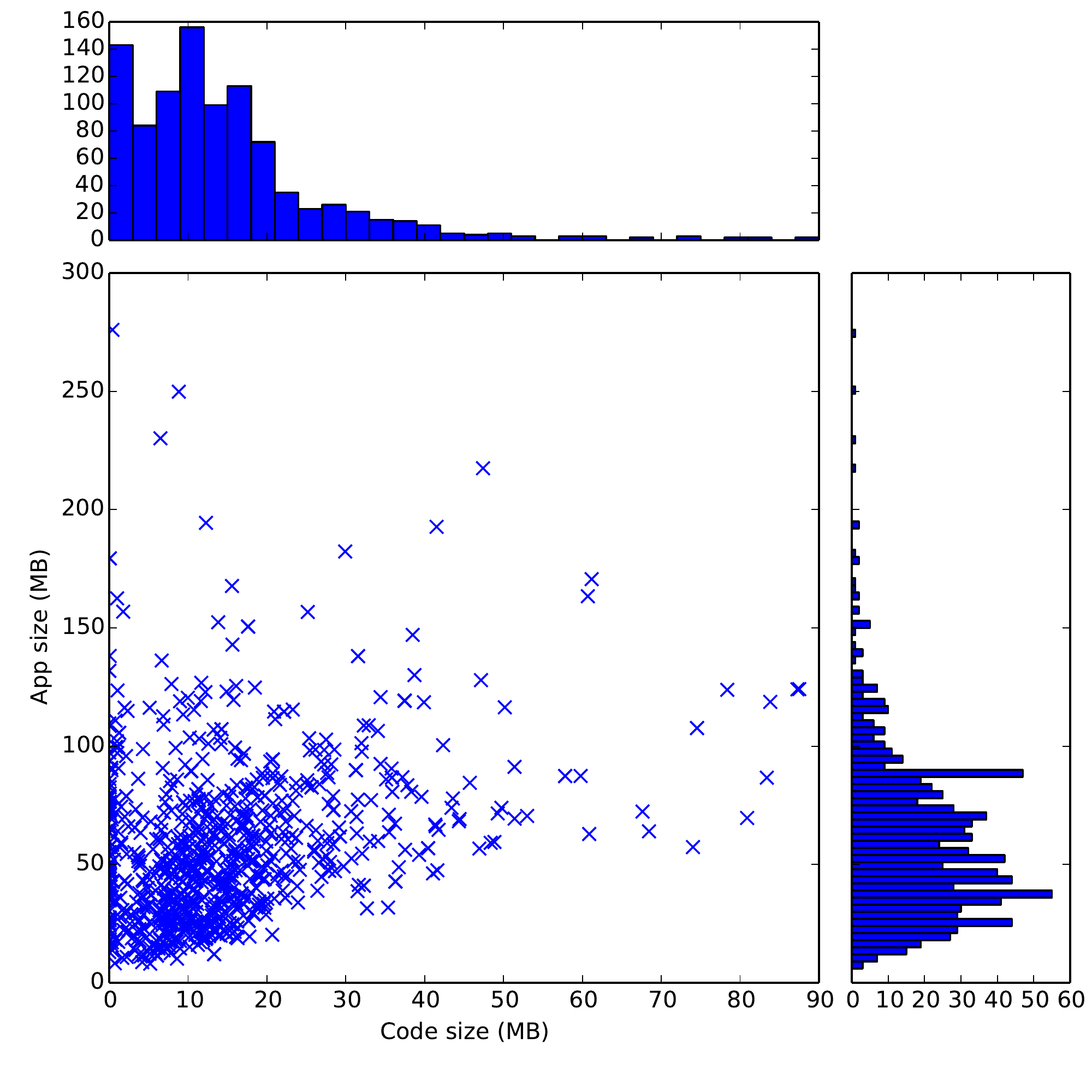}
	\caption{Statistics of the file size (app size and DEX code size) of all DL apps.}
	\label{fig:all_app}
	%\vspace{-5mm}
\end{figure} 

\subsection{\re{Failed App Analysis}}
\label{sec:failed}
\re{As shown in Figure~\ref{fig:failed_app}, we present the statistics of the file size characteristics of DL apps that cannot be run by Flowdroid. The results show that the app size distribution of the failed apps is relatively uniform, and the code size is mainly distributed in the range of 10-20MB. Besides, 23 of these apps are packed and 5 apps contain the package names filtered by Flowdroid by default.
We also conduct experiments to show the size statistics for all DL apps (See Figure~\ref{fig:all_app}). The average app size and DEX code size of all DL apps are 59.7MB and 14.4MB, respectively.}

%\clearpage
%\raggedend
%\input{revision}

% \section{Research Methods}

% \subsection{Part One}

% Lorem ipsum dolor sit amet, consectetur adipiscing elit. Morbi
% malesuada, quam in pulvinar varius, metus nunc fermentum urna, id
% sollicitudin purus odio sit amet enim. Aliquam ullamcorper eu ipsum
% vel mollis. Curabitur quis dictum nisl. Phasellus vel semper risus, et
% lacinia dolor. Integer ultricies commodo sem nec semper.

% \subsection{Part Two}

% Etiam commodo feugiat nisl pulvinar pellentesque. Etiam auctor sodales
% ligula, non varius nibh pulvinar semper. Suspendisse nec lectus non
% ipsum convallis congue hendrerit vitae sapien. Donec at laoreet
% eros. Vivamus non purus placerat, scelerisque diam eu, cursus
% ante. Etiam aliquam tortor auctor efficitur mattis.

% \section{Online Resources}

% Nam id fermentum dui. Suspendisse sagittis tortor a nulla mollis, in
% pulvinar ex pretium. Sed interdum orci quis metus euismod, et sagittis
% enim maximus. Vestibulum gravida massa ut felis suscipit
% congue. Quisque mattis elit a risus ultrices commodo venenatis eget
% dui. Etiam sagittis eleifend elementum.

% Nam interdum magna at lectus dignissim, ac dignissim lorem
% rhoncus. Maecenas eu arcu ac neque placerat aliquam. Nunc pulvinar
% massa et mattis lacinia.

\end{document}